\documentclass[prl,twocolumn,superscriptaddress,nobibnotes,a4paper]{revtex4-2}

\usepackage[pdftex]{graphicx}
\usepackage[pdftex]{epsfig}
\usepackage{amsmath}
\usepackage{amssymb}
\usepackage{amsfonts}
\usepackage{color}
\usepackage{wrapfig}
\usepackage{eucal}
\usepackage{hhline}
\usepackage{threeparttable}
\usepackage{supertabular}
\usepackage{multirow}
\usepackage{tabularx}
\usepackage{upgreek}
\usepackage{float}
\usepackage{siunitx}
\usepackage[normalem]{ulem}
\usepackage{lineno}

\usepackage[export]{adjustbox}

\usepackage{soul}

\usepackage[version=4]{mhchem} % for chemical formulas
\usepackage{array} % make column bigger (vertical)
\usepackage[colorlinks=true, allcolors=blue]{hyperref} %hyperrefs

\usepackage{comment}

\usepackage{booktabs,multirow,array}

\newcommand{\bra}[1]{\left\langle #1\right|}
\newcommand{\ket}[1]{\left|#1\right\rangle}

\setstcolor{red}

\graphicspath{ {./fig_pdf/} }
\usepackage{graphicx}

\begin{document}

\title{A unified framework for determining transition dipole polarization\\ in solid-state spin defects}

\author{Gaia Da Prato}
\affiliation{Kavli Institute of Nanoscience, Department of Quantum Nanoscience, Delft University of Technology, 2628CJ Delft, The Netherlands}

\author{Yong Yu}
\altaffiliation{Current address: Hefei National Laboratory, University of Science and Technology of China, Hefei 230088, China}
\affiliation{Kavli Institute of Nanoscience, Department of Quantum Nanoscience, Delft University of Technology, 2628CJ Delft, The Netherlands}

\author{Wolfgang Tittel}
\affiliation{Department of Applied Physics, University of Geneva, 1211 Geneva, Switzerland}
\affiliation{Constructor University Bremen, 28759 Bremen, Germany}

\author{Simon Gr\"oblacher}
\email{s.groeblacher@tudelft.nl}
\affiliation{Kavli Institute of Nanoscience, Department of Quantum Nanoscience, Delft University of Technology, 2628CJ Delft, The Netherlands}

\begin{abstract}
Spin–photon interfaces based on solid-state defects are key building blocks for scalable quantum networks and hybrid quantum platforms. Optimizing light--matter coupling in these systems requires precise knowledge of the optical transition dipole polarization, yet for many promising quantum emitters this quantity is hard to determine and therefore remains poorly characterized.
Here, we develop a framework for reconstructing electric transition dipole polarization in spin-1/2 solid-state defects directly from ensemble spectroscopy. The approach combines the response of photoluminescence spectra to magnetic field, optical polarization, and strain. Applied to erbium ions in silicon, a particularly challenging system containing multiple crystallographic subsites, the framework identifies strain-induced shifts as the origin of asymmetric ensemble spectra and enables simultaneous determination of the optical dipole polarization and strain--orbital coupling tensor. The resulting model predicts how cavity--ion coupling depends on crystallographic orientation and magnetic-field direction, which we verify using single erbium ions coupled to a nanophotonic cavity.
Together, these results establish a broadly applicable route for extracting microscopic properties of solid-state quantum emitters from ensemble spectroscopy and for engineering optimized spin--photon and spin--phonon interfaces.
\end{abstract}

\maketitle

\section*{Introduction}\label{sec_intro}
Spin–photon interfaces form a central component of emerging quantum technologies, enabling entanglement between stationary and flying qubits~\cite{Sangouard2011, Reiserer2015}, quantum network links~\cite{Kimble2008, Wehner2018}, and distributed sensing~\cite{Degen2017}. Among various quantum emitter platforms, including trapped neutral atoms~\cite{Sangouard2011, Reiserer2015}, ions~\cite{Duan2010}, and quantum dots~\cite{Lodahl2018}, solid-state defects~\cite{Awschalom2018, Castelletto2020, Ruf2021} have emerged as particularly promising owing to their scalability and ability to integrate with nanophotonic structures.
Optimal operation of such interfaces relies on how strongly the emitter interacts with light, which is ultimately governed by the transition dipole. Determining the dipole orientation is therefore essential for engineering efficiently coupled nanophotonic devices.
For some well-known systems such as selected color centers in diamond~\cite{Brown1995, Epstein2005, Rogers2014}, silicon~\cite{Durand2021}, and silicon carbide~\cite{Zhou2021}, the dipole properties are established through direct studies of spectrally isolated single emitters.
However, for many other emerging solid-state defects, including rare-earth ion-doped crystals (REIs), transition dipoles remain poorly characterized.

\begin{figure*}[htb!]
	\centering
    \includegraphics[]{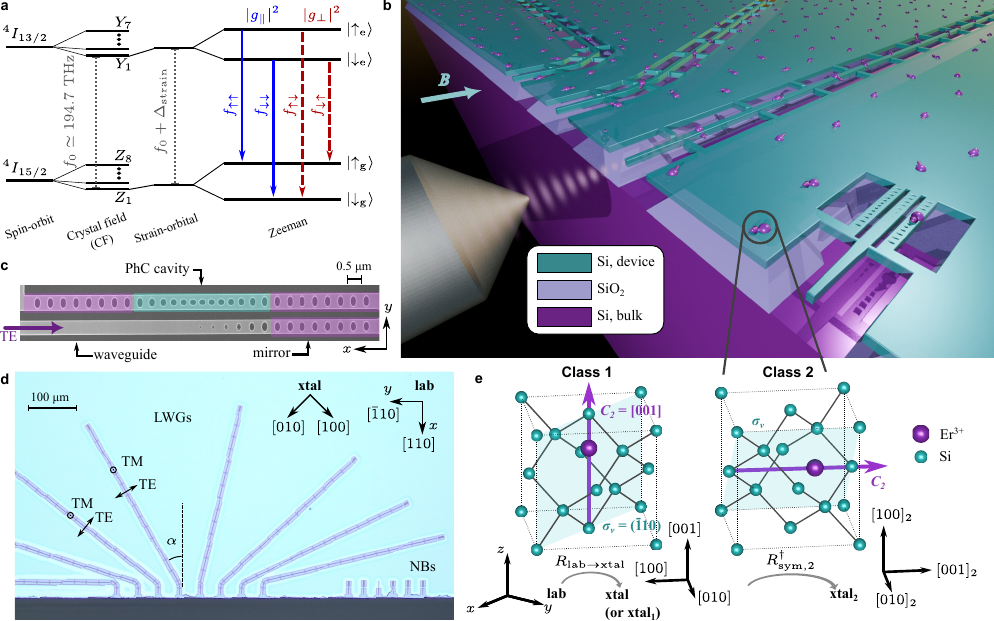}
	\caption{\textbf{Overview of the system.} 
		\textbf{a.}
		Energy levels of Er:Si. The diagram illustrates the hierarchy of interactions shaping the energy levels, including spin-orbit coupling, crystal field (CF) interaction, strain-orbital coupling, and Zeeman interaction. Spin-preserving (blue) and spin-flip (red) optical transitions have strengths $g_{\parallel}$ and $g_{\perp}$, respectively.
		\textbf{b.}
		Artist’s impression of the devices, showing a lensed fiber coupling light into a suspended long waveguide (LWG), and two photonic-crystal nanobeams (PhC NBs) with a central coupling waveguide (bottom right). The suspended structures are supported by tethers. Purple arrows represent individual Er$^{3+}$ defects in Si, whose quantization axes depend on the local $C_{2v}$ symmetry and on the applied magnetic field $\mathbf{B}$.
		\textbf{c.}
		Scanning electron microscope picture of a NB, evanescently coupled to a waveguide with a PhC mirror. 
        \textbf{d.}
        Optical microscope image of several devices, including LWGs and NBs. The orientations of the TE and TM modes are shown for waveguides fabricated at different angles~$\alpha$. The laboratory (lab) frame is conveniently defined along the chip axes, while the crystal (xtal) frame is aligned with the crystal axes.
        \textbf{e.}
        Example crystal structure consistent with the $C_{2v}$ point-group symmetry of the Er:Si site~A, illustrating two magnetic classes, together with their local $C_2$ axis and $\sigma_v$ mirror plane. Coordinate transformations between the laboratory, crystal, and class-specific frames ($\mathrm{xtal}_1$ and $\mathrm{xtal}_2$) are defined by $R_{\mathrm{lab}\rightarrow\mathrm{xtal}}$ and $R^{\dagger}_{\mathrm{sym},\,2}$. The local frame of class~1 ($\mathrm{xtal}_1$) is chosen to coincide with the crystal frame.
	}
	\label{fig1}
\end{figure*}

Due to the shielding of their 4f electrons, REIs~\cite{Liu2006, Thiel2011, Tittel2025} exhibit exceptionally long spin and optical coherence times~\cite{Zhong2015, Wang2025}, making them attractive for integrated spin--photon interfaces. 
Cavity-enhanced single-ion studies have demonstrated single photon emission~\cite{Zhong2018, Dibos2018, Yu2023, Gritsch2023}, single-shot spin readout~\cite{Kindem2020, Gritsch2025}, quantum nondemolition measurements~\cite{Raha2020}, indistinguishable photons from the same emitter~\cite{Ourari2023}, and entanglement protocols~\cite{Uysal2025, Ruskuc2025}, representing major advances toward quantum networking.
Among REI platforms, erbium in silicon (Er:Si) is particularly promising~\cite{Yin2013}, combining telecom C-band optical transitions and full compatibility with mature silicon photonics. Recent experiments have demonstrated distinct crystallographic sites with narrow linewidths~\cite{Gritsch2022}, single-shot spin readout~\cite{Gritsch2025}, and spin coherence times up to 1.2~ms~\cite{Berkman2025}. Notably, the so-called site A exhibits $C_{2v}$ symmetry~\cite{Holzapfel2024}, allowing electric-dipole contributions to the optical transition and resulting in an unusually short optical lifetime.
Yet key microscopic properties remain unknown, including the orientation of the optical transition dipole and the piezospectroscopic tensor~\cite{Kaplyanskii1967}, which characterizes strain--orbital coupling.
Knowledge of these quantities is essential for predicting and engineering efficient spin--photon interfaces. In particular, strain modifies the optical transitions and can potentially be exploited for their active tuning~\cite{Brevoord2025}.
In REIs, electric transition dipoles are difficult to calculate theoretically as they arise from higher-order orbital mixing beyond the crystal-field Hamiltonian~\cite{Liu2006}. Experimentally, weak spontaneous emission complicates direct single-ion studies and often makes ensemble measurements under specific crystallographic symmetry constraints the primary means of investigation~\cite{Luo2020, Kaloyeros2024, Becker2025}. Determining the piezospectroscopic tensor poses similar challenges and is further complicated by strain control and calibration, such that many studies report only phenomenological pressure sensitivities instead of a complete tensor characterization~\cite{Louchet2019, Zhang2019}.

Here, we overcome these limitations by developing a framework that unifies crystal symmetry, spin Hamiltonians, and strain--orbital coupling to reconstruct transition dipole polarization from spectroscopy. The framework is broadly applicable to optical transitions between effective spin-1/2 manifolds and is experimentally demonstrated for Er:Si.
We first reproduce ensemble photoluminescence (PL) spectra and infer the electric transition dipole polarization and piezospectroscopic tensor.
We then determine how the coupling of Er$^{3+}$ ions to a cavity depends on crystallographic subsite orientation and magnetic-field direction.
Lastly, we experimentally verify these predictions using single ions embedded in a nanocavity by identifying their orientation and performing field-dependent lifetime measurements.

\section*{System overview}\label{sec_system}

Figure~\ref{fig1}a sketches the optical transition around 194.7~THz between the $Z_1$ and $Y_1$ crystal-field levels of site~A of Er:Si.
Under a static magnetic field $\mathbf B$, the ground and excited states split into Kramers doublets $\ket{\downarrow_\mathrm{g}},\ket{\uparrow_\mathrm{g}}$ and $\ket{\downarrow_\mathrm{e}},\ket{\uparrow_\mathrm{e}}$, which can be described as effective spin-1/2 states governed by the spin Hamiltonians
\begin{equation}\label{eq:spin_hamiltonian}
\hat{H}_\mathrm{g(e)} (\mathbf{B})= \mu_{\mathrm B}\, \mathbf B \!\cdot\! g_\mathrm{g(e)} \!\cdot\! \hat{\mathbf S},
\end{equation}
where $\mu_\mathrm{B}$ is the Bohr magneton, $g_\mathrm{g(e)}$ is the ground-(excited-)state $g$-tensor, and $\hat{\mathbf S}=\tfrac12(\hat{\sigma}_x,\hat{\sigma}_y,\hat{\sigma}_z)$ is a vector containing the Pauli operators.

Optical transitions between the spin states are characterized by two dipole matrix elements: a spin-preserving dipole and a spin-flip dipole,
\begin{equation}\label{eq:dipole_def}
\mathbf d_{\parallel}=\bra{\downarrow_\mathrm{g}}\hat{\mathbf d}\ket{\downarrow_\mathrm{e}}, 
\qquad
\mathbf d_{\perp}=\bra{\downarrow_\mathrm{g}}\hat{\mathbf d}\ket{\uparrow_\mathrm{e}},
\end{equation}
where $\hat{\mathbf d}$ is the electric-dipole operator. The remaining matrix elements are fixed by the time-reversal relations within each Kramers pair and do not provide additional independent observables~\cite{Raha2020}. Given the short optical lifetime of site A, electric-dipole contributions are expected to dominate the transition, and magnetic-dipole terms are neglected~\cite{Gritsch2022, Holzapfel2024}.
Through PL measurements, we directly access the projection of the dipoles onto the optical field.
Spontaneous emission into a resonant electromagnetic mode with electric field $\mathbf E$ yields an intensity
\begin{equation}\label{eq:Fermi_golden}
I_{i} \propto |g_{i}|^{2}
          = \frac{|\mathbf d_{i}\cdot\mathbf E|^{2}}{\hbar^2}, \quad i \in \{\parallel,\perp\},
\end{equation}
where $g_{\parallel(\perp)}$ denotes the coupling strengths for the spin-preserving (spin-flip) transitions, and $\hbar$ is the reduced Planck constant.

As shown in Fig.~\ref{fig1}b, $\text{Er}^{3+}$ ions are implanted into a thin Si device layer (details in Appendix~\ref{app_sample_fabrication}). Owing to their weak emission, nanophotonic structures are used to guide and efficiently collect the emitted photons: suspended long waveguides (LWGs) for ensemble measurements and one-dimensional photonic-crystal nanobeams (PhC NBs, Fig.~\ref{fig1}c) for single-ion studies.
The LWGs support transverse electric (TE) and transverse magnetic (TM) modes. By fabricating waveguides at different angles $\alpha$ with respect to the silicon [110] crystal direction (Fig.~\ref{fig1}d), we control the mode polarization, described by the unit vector $\boldsymbol{\epsilon}(\alpha,m)$, where $m=\{\mathrm{TE},\mathrm{TM}\}$.
This allows us to project the transition dipoles onto different directions, and probe them with the measured PL.

Up to this point, we have described a single pair of dipoles $\mathbf d_{\parallel}$ and $\mathbf d_{\perp}$. In practice, however, these depend on both the magnetic field and the defect orientation.
$\mathbf B$ sets a preferred direction (the quantization axis) which influences the transitions, and as a consequence their dipole moments.
Importantly, the dipoles at any magnetic field orientation $\mathbf B$ are still fully determined by their values at a single reference orientation $\mathbf B_0$, namely $\mathbf d_{\parallel}(\mathbf B_0)$ and $\mathbf d_{\perp}(\mathbf B_0)$ (see Appendix~\ref{app_dipole_determination}).
In addition, the same Er$^{3+}$ defect can exist in multiple orientations within the Si lattice.
These different orientations correspond to twelve geometrically distinct subsites for site~A, which can be grouped into six so-called magnetically nonequivalent classes~\cite{Holzapfel2024, Luo2020} (see Appendix~\ref{app_magnetic_classes}).
All classes share identical intrinsic properties (e.g.\ $g$-tensors and symmetry axes) in their local frames. However, as illustrated in Fig.~\ref{fig1}e, they are oriented differently with respect to the laboratory frame.
As a consequence, external quantities—such as the magnetic field and the optical mode—couple differently to each class. This makes the transition dipoles and the spectral lines class dependent, even though the underlying defect is the same. In this sense, subsites belonging to different classes are nonequivalent.
We take $\mathbf d_{\parallel}(\mathbf B_0)$ and $\mathbf d_{\perp}(\mathbf B_0)$ to refer to class~1, without loss of generality, as they uniquely determine the dipoles of all other classes.

Therefore, the PL intensity is set by the magnetic class, external magnetic field, and optical mode, given the (to be determined) reference dipoles.
The full workflow is detailed in Appendix~\ref{app_workflow} and is broadly applicable to other spin-1/2 defects in solids.

\section*{Strain-induced asymmetry and dipole polarization from ensemble spectroscopy}\label{sec_ensemble}

We acquire PL excitation spectra with the sample placed in a dilution refrigerator ($T\approx10{-}30$ mK), ensuring that only the ground ${Z}_1$ levels are populated.
The emitted light is separated into TE and TM components prior to detection, allowing us to select the polarization $\boldsymbol{\epsilon}(\alpha,m)$ (see Appendix~\ref{app_exp_parameters}).
\begin{figure*}[htb!]
\centering
\includegraphics[]{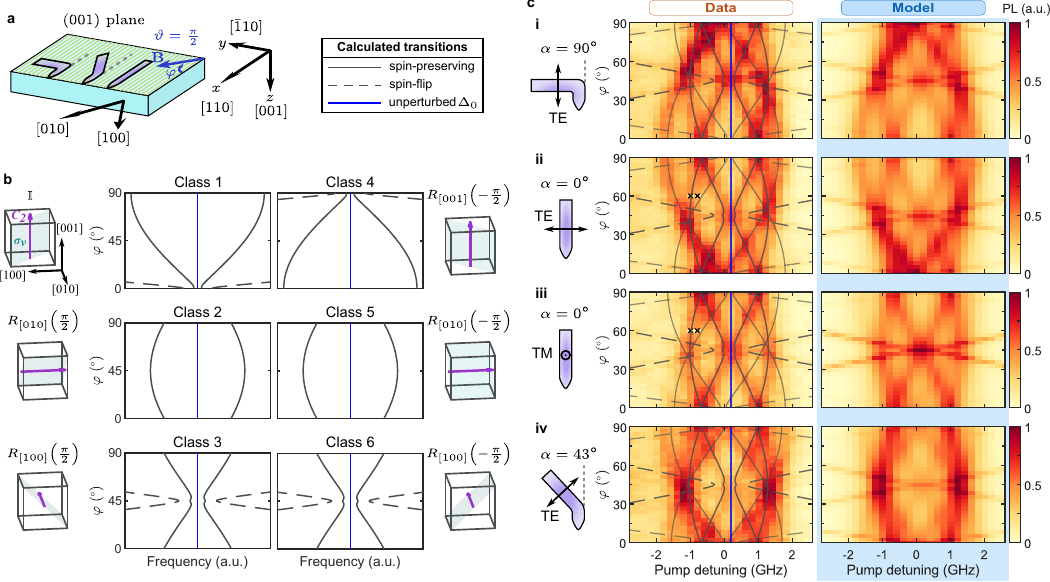}
\caption{\textbf{Magnetic field-dependent PL excitation spectra.}
\textbf{a.} The magnetic field (94~mT) is rotated in the $(001)$ plane, parameterized by $\varphi \in [0,\pi/2]$ with $\vartheta = \pi/2$.
\textbf{b.} Calculated field-dependent transition frequencies for each magnetic class, related by the indicated symmetry operations. Solid (dashed) lines denote spin-preserving (spin-flip) transitions, symmetric about the unperturbed frequency $\Delta_0$ (blue line).
\textbf{c.} PL excitation spectra measured on an ensemble (Data, left) and corresponding best-fit models (Model, right). Rows $i$–$iv$ correspond to different waveguide orientations $\alpha$ and selected optical modes (TE or TM), as illustrated. A subset of datasets is shown and the full datasets are reported in~\cite{SM}.
The laser detuning is referenced to the site~A zero-field transition at 1537.7607~nm~\cite{Gritsch2022}. Data and model are normalized by subtracting the background and dividing by the maximum signal. Crossed markers indicate data points excluded from the fit. The calculated transition frequencies from \textbf{b} are overlaid (gray lines), including a class-dependent strain-induced shift. 
The data are well reproduced by the model ($\chi_{\mathrm{red}}^2 = 7.30$, $R^2 = 0.71$; see Appendix~\ref{app_fit_goodness}), with residual deviations attributed to background PL from ions belonging to another site~\cite{SM}.
}
\label{fig2}
\end{figure*}
As shown in Fig.~\ref{fig2}a, the magnetic field $\mathbf{B}$ is rotated within the $(001)$ crystalline plane, parameterized by spherical angles $(\varphi,\vartheta)$, using a custom-built vector magnet~\cite{DaPrato2025}. This allows us to resolve the PL contributions from different magnetic classes and analyze their spectra individually.
Each spectrum, corresponding to a given magnetic class, contains four optical transitions arising from the level structure (Fig.~\ref{fig1}a).
In Fig.~\ref{fig2}b, these transitions trace out lines in angle--frequency space, indicating where bright PL is expected. Spin-preserving and spin-flip transitions form pairs that are symmetric about the zero-field transition frequency (blue line), as dictated by Eq.~\eqref{eq:spin_hamiltonian}.
Each panel corresponds to a different class, whose orientations are related by symmetry operations of the host lattice (as illustrated in the diagrams), giving rise to distinct patterns.
These patterns can be understood by considering the geometrical projection of the classes onto the $(001)$ plane, to which $\mathbf{B}$ is confined. The spectra of classes 1 and 4 are related by a reflection about $\varphi = 45^\circ$, consistent with their $90^\circ$ rotational relation about the [001] axis. Classes 2 and 5, as well as classes 3 and 6, instead become indistinguishable in this configuration, yielding identical spectra that are symmetric about $\varphi = 45^\circ$.

A selection of measured PL colormaps from an ensemble of ions in LWGs is shown in Fig.~\ref{fig2}c (Data panels). The complete datasets, including additional optical modes and $\mathbf{B}$-field directions, are reported in~\cite{SM}.
In an ensemble, all classes contribute to the PL. Assuming they are generated with equal probability during implantation, the expected spectrum is an equal superposition of the patterns in Fig.~\ref{fig2}b symmetric about a common center frequency.
However, the experimental PL colormaps in Fig.~\ref{fig2}c deviate from this behavior:\ while multiple spectral lines corresponding to different classes are observed, the patterns are not centered around the same frequency, resulting in an overall asymmetric map.
This deviation cannot be explained by the spin Hamiltonian alone. Instead, it reveals the presence of class-dependent strain-orbital coupling.
For each class, strain induces a common shift of the four optical transitions (Fig.~\ref{fig1}a). We model this effect using a piezospectroscopic tensor $A$~\cite{Kaplyanskii1967, Buzzi2025}, such that the frequency shift is
\begin{equation}\label{eq:strain_coupling}
\Delta_{\mathrm{strain}} = \sum_{i,j} A_{ij}\,\sigma_{ij},
\end{equation}
where $\sigma$ is the stress tensor, and $A = A_{\mathrm{e}} - A_{\mathrm{g}}$ is the difference between the excited- and ground-state coupling tensors.
As $A$ is an intrinsic property of the defect, while $\sigma$ is defined in the laboratory frame, their relative orientation varies between subsites, leading to class-dependent $\Delta_{\mathrm{strain}}$ and thus the observed asymmetry.
In our devices, strain arises primarily from inherent strain in the thin-film silicon layer from fabrication, as well as thermal contraction of the suspended waveguides upon cooling from room temperature to cryogenic temperatures. We estimate the average strain tensor from simulations and infer the corresponding stress~\cite{SM}, enabling extraction of the piezospectroscopic tensor $A$ from fits to the transition frequencies (overlaid lines in Fig.~\ref{fig2}c). While the strain fields induced by thermal contraction are relatively small, they already permit robust determination of the dominant strain response. The largest-magnitude eigenvalue is $-7.8\pm0.3~\mathrm{Hz/Pa}$, which we take as the characteristic coupling scale (see Appendix~\ref{app_strain_orbital_coupling} for the tensor components).
This is comparable to previously reported strain sensitivities for Er:Si~\cite{Zhang2019}, although those measurements did not resolve the crystallographic site or orientation and thus provided only a scalar projection of the tensor. It is smaller than typical values reported for REIs in oxide or fluoride hosts~\cite{Louchet2019}, consistent with the large stiffness, high symmetry, and predominantly covalent nature of the Si lattice.
This interpretation is further supported by the absence of such asymmetry in previous measurements on non-suspended structures~\cite{Holzapfel2024}, which experience significantly lower strain than our waveguides.

Having determined the transition frequencies, we now turn to the corresponding PL intensity, governed by the interaction of the transition dipole with the optical mode (see Eq.~\eqref{eq:Fermi_golden}). In Fig.~\ref{fig2}c (Data panels), spectral lines associated with different classes exhibit strongly varying intensities within a given colormap, indicating distinct dipole orientations.
Certain lines (e.g.\ class 1 in Fig.~\ref{fig2}c(i)) fall below the noise level due to weak coupling to the optical field, but reappear for other waveguide orientations or modes (e.g., in Fig.~\ref{fig2}c(ii, iv)), reflecting the strong dependence of the PL intensity on the optical configuration.
The $\alpha = 0^\circ$ and $\alpha = 90^\circ$ cases (Fig.~\ref{fig2}c(i, ii)) are related by exchanging the $x$ and $y$ components of the magnetic field, leading to corresponding symmetry relations in the PL. In contrast, when the optical mode is invariant under this interchange (Fig.~\ref{fig2}c(iii, iv)), the colormaps are symmetric about $\varphi = 45^\circ$.
The unknown reference dipoles fully determine the PL intensity for all classes. We extract their polarization through a global fit of all colormaps, yielding the results shown in Fig.~\ref{fig2}c (Model panels), which are in good agreement with the data. Details of the fitting procedure and goodness-of-fit metrics are reported in Appendix~\ref{app_fit_goodness} and~\cite{SM}.

\section*{Cavity-ion coupling characterization}

\begin{figure}[htb!]
	\centering
	\includegraphics[]{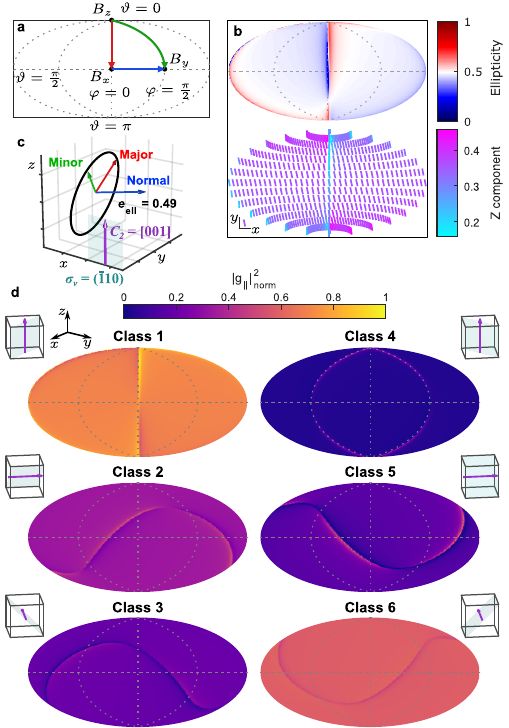}
    \caption{\textbf{Calculated dipole properties as a function of magnetic-field orientation for the spin-preserving transition.}
    \textbf{a.}
    Hammer projection of the magnetic-field orientation in the laboratory frame. Red, blue, and green arrows indicate field sweeps in the $zx$-, $xy$-, and $zy$-planes, respectively. Panels \textbf{b-d.} show quantities computed from the fitted reference dipoles as a function of field orientation.
    \textbf{b.}
    Dipole polarization properties for a class~1 ion. Top: ellipticity $e_{\mathrm{ell}}$, ranging from 0 (linear) to 1 (circular). Bottom: vector-field map of the major-axis direction projected onto the $xy$-plane; the colormap encodes its $z$ component, which is positive by normalization.
    \textbf{c.}
    Mean polarization ellipse ($e_{\mathrm{ell}} = 0.49$) obtained by averaging the major, minor, and normal axes over all magnetic-field orientations, excluding a $10^\circ$ region around the $zx$-plane to capture typical behavior. Red, green, and blue arrows indicate the mean major, minor, and normal axes, with directions $[-0.18, 0.92, 0.35]$, $[-0.10, -0.37, 0.92]$, and $[0.97, 0.15, 0.17]$ in the laboratory frame.
    \textbf{d.}
    Normalized squared coupling strength $|g_{\parallel}|_\mathrm{norm}^2$ (Eq.~\eqref{eq:norm_coupling_strength}) of the dipole to a NB optical mode polarized along $y$.
    Each panel corresponds to the specified magnetic class, oriented as shown in the schematics alongside the map.
    }
	\label{fig3}
\end{figure}

\begin{figure*}[htb!]
\centering
\includegraphics[]{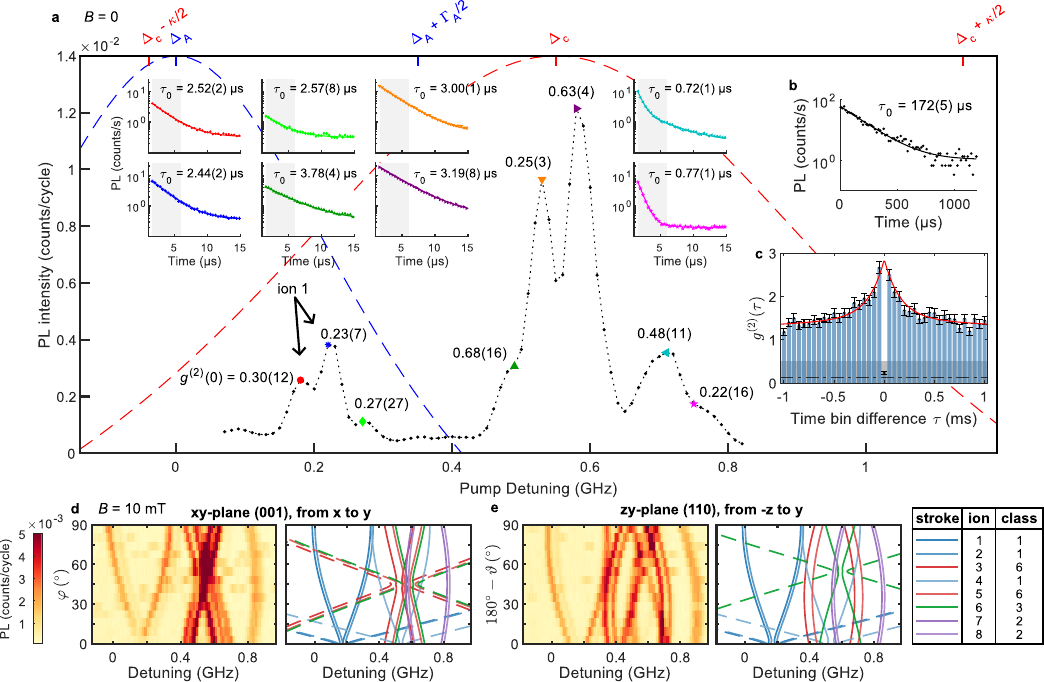}
\caption{\textbf{PL measurements of single ions coupled to a PhC cavity.}
\textbf{a.} 
PL excitation spectrum at zero magnetic field. Frequency detunings are referenced to the center ($\Delta_{\mathrm{A}}$, 1537.7607~nm) of the site~A inhomogeneous broadening indicated by the blue dashed profile of linewidth $\Gamma_{\mathrm A}$. The cavity resonance (red dashed Lorentzian profile; center $\Delta_{\mathrm c}$, linewidth $\kappa$) is positioned on the tail of the inhomogeneous broadening, enabling spectral resolution of individual ions.
Spectral peaks are highlighted with colored markers (error bars smaller than the markers) and labeled with the corresponding $g^{(2)}(0)$ values. Insets show time-resolved PL traces of matching color, measured from the end of the excitation pulse; shaded regions indicate the integration windows used to extract the PL intensities. Solid curves are single- or double-exponential fits~\cite{SM}, with the short lifetime component $\tau_0$ indicated. The reduced lifetimes are compared with the bulk lifetime of \SI{172(5)}{\micro\second}, extracted from time-resolved PL of an ensemble of ions in a LWG (panel~\textbf{b}).
The spin-preserving transitions of the selected ion~1, split by a stray magnetic field, are indicated by arrows in panel~\textbf{a}.
A longer acquisition of the high-frequency transition of ion~1 provides a more accurate estimate of $g^{(2)}(\tau)$ (panel~\textbf{c}), resulting in $g^{(2)}(0)=0.24(4)$. The dash-dotted line indicates the estimated noise contribution ($g^{(2)}_{\mathrm{noise}}(0)=0.13$), yielding an effective value $g_{\mathrm{eff}}^{(2)}(0)=0.11(4)$. Bunching at nonzero delay $\tau$ is fitted with a double-exponential model (red trace)~\cite{SM}.
\textbf{d, e.} Left: measured PL colormaps versus pump detuning and magnetic-field orientation $(\varphi,\vartheta)$ at fixed magnitude of 10~mT. In \textbf{d}, $\vartheta=\pi/2$ and $\varphi$ is scanned from 0 to $\pi/2$; in \textbf{e}, $\varphi=\pi/2$ and $\vartheta$ is scanned from $\pi$ to $\pi/2$. Right: predicted spectral lines based on the magnetic classes inferred from the data in the left panels, with color indicating magnetic class and shade distinguishing individual ions; solid (dashed) lines indicate spin-preserving (flip) transitions.
}
\label{fig4}
\end{figure*}

Starting from the fitted reference dipoles $\mathbf d_{\parallel}(\mathbf B_0)$ and $\mathbf d_{\perp}(\mathbf B_0)$, we compute the polarization properties for arbitrary magnetic-field orientations $\mathbf{B}(\varphi,\vartheta)$, represented using a Hammer projection (Fig.~\ref{fig3}a). 
We focus on the spin-preserving transition, which is typically the strongest (see~\cite{SM} for the spin-flip case).
Fig.~\ref{fig3}b shows quantities defining the dipole polarization for a representative class (class~1), as the behavior is identical for all classes up to a frame transformation~\cite{SM}. Both the ellipticity (top) and the major axis (bottom) vary only weakly over the sphere, except for magnetic fields near the $xz$-plane.
This behavior originates from the strong anisotropy of the ground- and excited-state $g$-tensors, whose principal axis lies along $y$, pinning the quantization axes $g_{\mathrm{g(e)}} \cdot \mathbf{B}$ close to this direction for most field orientations. 
It is only when the field approaches the plane perpendicular to $y$ that the weaker $g$-tensor components compete, allowing the quantization axis—and therefore the transition dipoles—to tilt.
The mean polarization ellipse is shown in Fig.~\ref{fig3}c, with ellipticity 0.49, the major axis oriented close to $y$, and the normal approximately along $x$. This suggests that a class~1 ion couples nearly optimally to a NB TE mode polarized along $y$, as is the case for our devices.
For each magnetic class, we then predict the squared normalized coupling strength
\begin{equation} \label{eq:norm_coupling_strength}
 |g_{\parallel}|_{\mathrm{norm}}^2 = \frac{|g_{\parallel}|^2}{|\mathbf{E}|^2|\mathbf{d}_{\parallel}|^2}
\end{equation}
to a NB cavity. 
As shown in Fig.~\ref{fig3}d, the coupling is nearly independent of the magnetic-field orientation and largest for class~1, consistent with the polarization properties discussed above. It is smallest for class~4, whose dipole is related to that of class~1 (Fig.~\ref{fig3}c) by an interchange of $|x|$ and $|y|$, reducing its overlap with the cavity field; the remaining classes lie in between.
The nearly constant coupling strength contrasts with Ref.~\cite{Raha2020}, where the cyclicity exhibits a strong magnetic-field dependence. This difference originates from the misalignment of $g_{\mathrm{g}}$ and $g_{\mathrm{e}}$, which introduces competing principal axes~\cite{SM}.

Next, we experimentally verify our model predictions on single ions coupled to a NB PhC cavity (Fig.~\ref{fig1}c). The cavity linewidth (FWHM) is $\kappa/2\pi = 1.2$~GHz, corresponding to $Q = 1.6\times10^5$. The resonance frequency $\Delta_{\mathrm{c}}$ is gas-tuned by $\sim$1~nm to lie near the center $\Delta_\mathrm{A}$ of the site~A inhomogeneous broadening~\cite{SM}.
Fig.~\ref{fig4}a shows the cavity-enhanced PL excitation spectrum at zero magnetic field. For the highlighted local maxima, the zero-delay second-order correlation $g^{(2)}(0)$ (reported next to the markers) is well below~1, indicating non-classical few-photon emission, and in some cases below~0.5, confirming single-photon emission~\cite{Loudon2000}. The corresponding time-resolved PL count rates (insets) exhibit single- or double-exponential decays, as expected from the number of emitters~\cite{SM}. Comparing the reduced lifetimes with the bulk value (Fig.~\ref{fig4}b) yields Purcell factors between~45 and~239, in line with the state of the art for Er:Si~\cite{Gritsch2025}.

To identify the magnetic class of each ion, we record PL excitation spectra for different magnetic-field orientations in the $(001)$ plane (Fig.~\ref{fig4}d, left panel).
As discussed in Fig.~\ref{fig2}b, each class exhibits characteristic field-dependent transition frequencies. However, in the $(001)$ plane, classes (2,5) and (3,6) produce identical patterns.
By combining these measurements with rotations in a second plane—$(110)$ in our case (Fig.~\ref{fig4}e, left panel)—we obtain sufficient information for unique classification (see Appendix~\ref{app_class_identification}).
Using this approach, we identify eight ions and assign their magnetic class. The corresponding theoretical transitions are shown in Fig.~\ref{fig4}d,e (right panels), color-coded by class.
Three ions belong to class~1—the class predicted to couple most strongly to the NB—while none correspond to class~4, which couples most weakly. Class~5 is also absent, likely due to the limited number of ions inside the cavity.

Among the identified emitters, we focus on a spectrally well-isolated ion (“ion~1”), belonging to class~1 and therefore ideally suited for a detailed study of the cavity–ion coupling. The peaks indicated by the arrows in Fig.~\ref{fig4}a correspond to its spin-preserving transitions, split by a small residual stray magnetic field.
To improve the accuracy, we repeat the $g^{(2)}(\tau)$ measurement for the higher-energy peak with longer integration time (Fig.~\ref{fig4}c). After subtracting the noise contribution, we obtain an effective value $g_{\mathrm{eff}}^{(2)}(0)=0.11\pm0.04$, confirming its single-ion nature.
The remaining deviation from zero likely arises from ions in the coupling waveguide and in the NB that are not Purcell-enhanced. Photon bunching at non-zero delays is consistent with spin-flip and spectral-diffusion dynamics~\cite{Gritsch2023, Bowness2025, SM}.

\begin{figure}[htb!]
\centering
\includegraphics[]{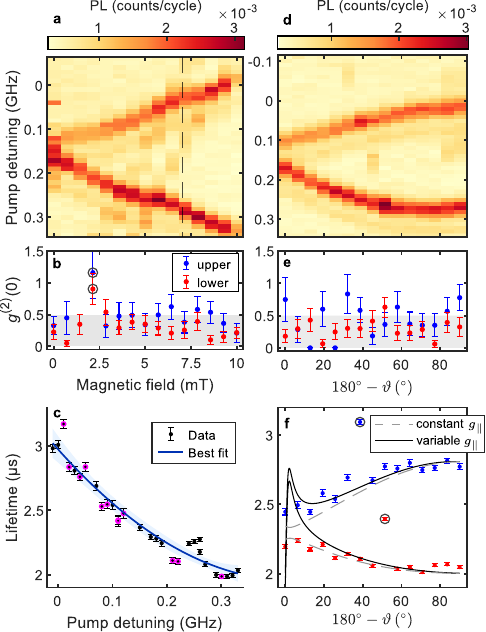}
\caption{\textbf{Magnetic-field-dependent lifetime measurements on ion~1.} 
\textbf{a.}~PL excitation spectra versus magnetic-field magnitude applied along $y$, with pump detuning referenced to 1537.7607~nm. The prominent upper and lower branches correspond to the two spin-preserving transitions of ion~1. Faint replicas at slightly higher frequency are attributed to another ion, while weak diagonal lines from spin-flip transitions of nearby ions cross the branches of ion~1. The dashed line marks the magnetic field used for the angular scan in panel~\textbf{d}.
\textbf{b.}~Second-order correlation $g^{(2)}(0)$ versus magnetic-field magnitude for the upper (blue) and lower (red) branches in \textbf{a}. Circled points indicate measurements affected by detector noise (one outlier with $g^{(2)}(0)=2.2$ at 1.4~mT is omitted). The shaded region below 0.5 marks the single-photon regime.
\textbf{c.}~Frequency-dependent lifetimes, obtained from exponential fits to time-resolved PL of the two branches in \textbf{a}. The blue curve shows the best fit to Eq.~\eqref{eq:enhanced_emission_B_field}, with the shaded region indicating the 95\% confidence interval; the fitted parameters are $g_{\parallel}/2\pi = 4.92 \pm 0.45$~MHz, $\kappa/2\pi = 1.19 \pm 0.16$~GHz, and cavity detuning $\Delta_{\mathrm{c}}/2\pi = 0.43 \pm 0.08$~GHz. Circled points correspond to measurements with $g^{(2)}(0)\ge 0.5$.
\textbf{d.}~PL excitation spectra versus magnetic-field orientation $\vartheta$ in the $zy$ plane (from $-z$ to $y$).
\textbf{e.}~Same as \textbf{b}, for the data in \textbf{d}.
\textbf{f.}~Fitted lifetimes versus $\vartheta$ for the upper (blue) and lower (red) branches in \textbf{d}. Two models based on Eq.~\eqref{eq:enhanced_emission_B_field} are shown:\ the gray dashed curve assumes a constant $g_{\parallel}$ (from \textbf{c}), while the black solid curve uses the angular dependence $g_{\parallel}(\varphi,\vartheta)$ from Fig.~\ref{fig3}d. Outliers due to spectral crossings with other ions are circled.
}
\label{fig5}
\end{figure}

We investigate the dependence of the coupling strength $g_{\parallel}$ on the external magnetic field to experimentally test the predictions in Fig.~\ref{fig3}d.
The coupling is probed via the Purcell-enhanced emission rate~\cite{Reiserer2015},
\begin{equation}
\label{eq:enhanced_emission_B_field}
\Gamma_{\text{P}}(B,\varphi,\vartheta) =
\frac{4|g_{\parallel}(\varphi,\vartheta)|^2}{\kappa}
\frac{1}{1 + \left(\frac{2\Delta(B,\varphi,\vartheta)}{\kappa}\right)^2},
\end{equation}
where $\Delta$ is the ion-cavity detuning.
To isolate $g_{\parallel}$, we first fix the field orientation to $\mathbf{B} \parallel y$ and vary its magnitude, which tunes $\Delta$ while leaving $g_{\parallel}$ unchanged. Fig.~\ref{fig5}a shows the resulting PL spectra, where the two bright branches correspond to the spin-preserving transitions of ion~1. For each field value, we report $g^{(2)}(0)$ in Fig.~\ref{fig5}b to verify that the single-ion character is preserved.
The frequency-dependent lifetimes are shown in Fig.~\ref{fig5}c, obtained from exponential fits to the time-resolved PL of each branch (see Appendix~\ref{app_lifetime_data_processing}).
Fitting these data with Eq.~\eqref{eq:enhanced_emission_B_field} yields $g_{\parallel}/2\pi = 4.92 \pm 0.45$~MHz and cavity parameters ($\kappa$, $\Delta_{\mathrm{c}}$) consistent with independent characterizations~\cite{SM}. This confirms that the lifetime variation is well described by detuning-dependent Purcell enhancement. Small deviations are attributed to uncertainties in the laser frequency~\cite{SM} and to spectral crossings with spin-flip transitions of nearby ions or partial overlap with a weaker adjacent ion (Fig.~\ref{fig5}a), as supported by the elevated $g^{(2)}(0)$ of several points (Fig.~\ref{fig5}b).

We then investigate the angular dependence of the coupling by fixing the field magnitude to 7~mT (dashed vertical line in Fig.~\ref{fig5}a) and rotating its direction in the $zy$-plane, as shown in Fig.~\ref{fig5}d,e. The extracted lifetimes as a function of magnetic-field orientation (Fig.~\ref{fig5}f) are compared with two models based on Eq.~\eqref{eq:enhanced_emission_B_field}:\ one assuming a constant $g_{\parallel}$ equal to the fitted value above, and one incorporating the angular dependence predicted in Fig.~\ref{fig3}d, normalized to the same fitted value.
Because the dipole orientation varies only weakly with magnetic-field direction (Fig.~\ref{fig3}), the two models are nearly identical over most angles, with only a small deviation predicted near the $z$ axis ($\vartheta=\pi$). 
The data agree well with both models over most angles. However, experimental uncertainties in field alignment and perturbations from nearby ions limit our ability to resolve the narrow feature near $\vartheta=\pi$~\cite{SM}. Consistently, the two circled outliers in Fig.~\ref{fig5}f occur at angles where ion-ion spectral crossings are present (Fig.~\ref{fig5}d), indicating that nearby emitters can modify the measured lifetime and obscure any fine angular dependence.

\section*{Discussion}

We establish a theoretical and experimental framework to reconstruct optical transition dipole polarization in solid-state spin defects directly from ensemble spectroscopy. The approach combines magnetic-field-, optical polarization-, and strain-dependent spectroscopy, while incorporating the distinct response of multiple crystallographic subsites when present. This overcomes limitations of existing methods that rely on spectrally isolated single defects or specific symmetry constraints~\cite{Epstein2005, Rogers2014, Durand2021, Zhou2021, Kaloyeros2024, Becker2025}. In its present form, the framework applies broadly to spin-1/2 defects with predominantly electric-dipole optical transitions that can be described within a spin Hamiltonian formalism. More generally, the underlying formalism can be extended to higher-spin manifolds and to emitters with mixed electric- and magnetic-dipole character~\cite{Raha2020}.

We demonstrate that class-dependent asymmetries in ensemble spectra provide a direct spectroscopic signature of strain--orbital coupling. 
Applying the framework to site A of Er:Si yields, to the best of our knowledge, the first quantitative determination of the piezospectroscopic tensor for erbium in any host material.
Unlike existing approaches for REIs that typically measure only phenomenological stress sensitivities~\cite{Louchet2019, Zhang2019}, our method provides access to the underlying tensorial form.
Future experiments with controlled strain fields, for example using MEMS devices~\cite{Buzzi2025, Brevoord2025}, could further refine the tensor components and enable deterministic strain engineering of optical transitions for spectral matching of remote emitters.

The present results also provide new insight into spin--photon interfaces based on Er:Si.
Using a photonic crystal cavity, we probe individual ions with optical lifetimes as short as \SI{0.72(1)}{\micro\second}, representing more than a $10\%$ improvement over the best previously reported values for erbium emitters~\cite{Gritsch2025}.
For each ion, we demonstrate for the first time identification of its class through magnetic-field rotations, an essential ingredient for predicting coupling to external perturbations.
We further reveal that the tunability of optical dipoles under magnetic-field rotation is fundamentally determined by the form of the ground- and excited-state $g$-tensors. In our system, the strong anisotropy of these tensors results in a cavity--ion coupling strength that is only weakly tunable, as confirmed by single-ion lifetime measurements.
Optimization of the coupling is therefore more naturally achieved through engineering cavity field polarization~\cite{Hallett2022, Zhu2025} to match the elliptical transition dipole rather than by tuning the dipole itself.
Resolving the sharp lifetime variations predicted near specific field directions would provide a direct test of the relation between $g$-tensor anisotropy and dipole orientation.
This regime can be achieved with improved magnetic-field alignment and enhanced spectral isolation of individual emitters, for example through reduced implantation densities, deterministic ion placement near the cavity field maximum, or larger magnetic fields to increase Zeeman splittings along weak-$g$ directions.
Beyond Er:Si, spin defects with different crystal symmetries and $g$-tensor forms may exhibit substantially stronger dipole tunability~\cite{SM}, enabling control of cavity--ion coupling through magnetic-field orientation~\cite{Raha2020}.

Furthermore, our work establishes Er:Si nanobeams as a promising hybrid quantum platform.
The Gigahertz mechanical modes supported by our nanobeams~\cite{Hong2017} provide a route for translating the static strain coupling characterized here into coherent spin--phonon interactions in the quantum regime~\cite{Wang2020}.
Developing hybrid spin--optomechanical platforms based on REIs could thereby open new opportunities for nonlinear quantum optomechanics~\cite{Aspelmeyer2014, Barzanjeh2022}, enabling the engineering of non-Gaussian mechanical states~\cite{Lachman2022, Rakhubovsky2024}.

\textbf{Acknowledgments.}
We would like to thank Viatcheslav Dobrovitski, Xiong Yao, Luca Mastrangelo, Annalise Lennon, and Emanuele Urbinati for their help, and Adrian Holz\"apfel for discussions on the theoretical model. This work is financially supported by the European Research Council (ERC CoG Q-ECHOS, 101001005) and by the Netherlands Organization for Scientific Research (NWO) as part of Vici (VI.C.222.024), Quantum Delta NL Quantum Technology (NGF.1623.23.009) and Klein (OCENW.KLEIN.555) grants. Y.Y. gratefully acknowledges funding from the European Union under a Marie Skłodowska-Curie fellowship. W.T. acknowledges funding through the Netherlands Organization for Scientific Research, the European Union’s Horizon 2020 Research and Innovation Program under Grant Agreement No.\ 820445, the Quantum Internet Alliance, and the Swiss State Secretariat for Education, Research and Innovation (SERI) under Contract Number UeM019-3.\\

\textbf{Author contributions.}
G.D.P., Y.Y., W.T., and S.G. conceived the research project. G.D.P. and Y.Y. designed and fabricated the devices. G.D.P. carried out the measurements with assistance from Y.Y.. G.D.P. developed the theoretical model, performed the simulations, and analyzed the data. S.G. and W.T. supervised the work. G.D.P. wrote the manuscript with input from all the authors.

\appendix

\section{Sample fabrication}\label{app_sample_fabrication}
The devices are fabricated from silicon-on-insulator (SOI) with $\langle 100 \rangle$ orientation, consisting of a 250~nm-thick Si device layer and a \SI{3}{\micro\meter} buried oxide layer. Erbium ions are implanted at room temperature with an energy of 350~keV and a dose of $1\times10^{12}$~ions/cm$^2$ (ITC, Japan). Following implantation, the sample undergoes rapid thermal annealing at 500~$^\circ$C for 1~min in a nitrogen atmosphere, with ramp-up and ramp-down times of 27~s and 200~s, respectively.  

After annealing, the photonic devices are patterned by electron-beam lithography and reactive ion etching of the Si layer, followed by selective removal of the buried oxide in hydrofluoric acid to release the suspended structures. Additional details are provided in~\cite{SM}.

\section{Energy levels}
We refer to the energy-level scheme shown in Fig.~\ref{fig1}a. We consider site~A of Er$^{3+}$-doped Si~\cite{Gritsch2022}. The relevant optical transitions occur between the $^4I_{15/2}$ ground manifold and the $^4I_{13/2}$ excited manifold, which arise from spin-orbit coupling. In free-space ions, these manifolds are 16-fold and 14-fold degenerate, respectively~\cite{Liu2006}. For Kramers ions in a solid-state host, the degeneracy is partially lifted by the crystal-field (CF) Hamiltonian, yielding eight sublevels in the ground state ($Z_1, \ldots, Z_8$) and seven in the excited state ($Y_1, \ldots, Y_7$). We focus on the zero-phonon-line (ZPL) optical transition between the lowest double-degenerate states, $Z_1$ and $Y_1$.
When an external magnetic field is applied, time-reversal symmetry is broken, and each doublet splits into spin-up and spin-down components, forming an effective spin-1/2 system. Strain perturbs the crystal lattice and therefore the crystal-field Hamiltonian, leading to energy shifts that differ for the ground and excited states and consequently modify the optical transition energies.

\section{$g$-tensors in the crystal and laboratory bases}
The structure of the $g$-tensors is determined by the local $C_{2v}$ point group
symmetry of Er:Si site A.
In the crystal basis defined by the Si lattice directions (see Fig.~\ref{fig1}d,e), the measured
tensors are~\cite{Holzapfel2024}:
\begin{equation}
g_{\mathrm{g},\mathrm{xtal}} = 
\begin{pmatrix}
8.5 & 8.1 & 0\\
8.1 & 8.5 & 0\\
0 & 0 & 0.58
\end{pmatrix},
\end{equation}
\begin{equation}
g_{\mathrm{e},\mathrm{xtal}} = 
\begin{pmatrix}
6.94 & 6.72 & 0\\
6.72 & 6.94 & 0\\
0 & 0 & 0.24
\end{pmatrix}.
\end{equation}

Because both tensors arise from Kramers doublets at the same $C_{2v}$ site,
they share a common set of principal axes and can therefore be simultaneously
diagonalized.
We define the resulting principal axis frame as the laboratory basis.  
In this basis, the tensors take the diagonal form
\begin{equation}
g_{\mathrm{g},\mathrm{lab}} = 
\begin{pmatrix}
0.40 & 0 & 0\\
0 & 16.6 & 0\\
0 & 0 & 0.58
\end{pmatrix},
\end{equation}
\begin{equation}
g_{\mathrm{e},\mathrm{lab}} = 
\begin{pmatrix}
0.22 & 0 & 0\\
0 & 13.66 & 0\\
0 & 0 & 0.24
\end{pmatrix}.
\end{equation}
The rotation matrix transforming vectors from the laboratory to the crystal basis is:
\begin{equation}
R_{\mathrm{lab}\rightarrow\mathrm{xtal}} =
\begin{pmatrix}
1/\sqrt{2} & -1/\sqrt{2} & 0\\
1/\sqrt{2} &  1/\sqrt{2} & 0\\
0 & 0 & 1
\end{pmatrix}.
\end{equation}

\section{Magnetic classes}\label{app_magnetic_classes}
At zero magnetic field, site~A contains twelve geometrically distinct subsites that are related by symmetry operations of the Si lattice ($O_h$ point group) which are not elements of the local $C_{2v}$ site symmetry. Despite their different orientations, these subsites share the same crystal field Hamiltonian and therefore have identical eigenfrequencies.
A magnetic field of generic orientation lifts all spatial symmetries of the Hamiltonian through the Zeeman interaction, preserving only inversion. Consequently, the twelve subsites group into six magnetic classes, each formed by a pair of inversion-related subsites. Within each class the eigenfrequencies remain identical, while different classes are generally nonequivalent and exhibit distinct spectra.
The spatial symmetry operations used to generate the six magnetic classes are
expressed in the crystal basis as:
\begin{equation}
R_{\mathrm{sym},1} \equiv \mathbb{I} = 
\begin{pmatrix}
1 & 0 & 0\\
0 & 1 & 0\\
0 & 0 & 1
\end{pmatrix},
\end{equation}

\begin{equation}
R_{\mathrm{sym},2} \equiv R_{[010]}\!\left(\frac{\pi}{2}\right) = 
\begin{pmatrix}
0 & 0 & -1\\
0 & 1 & 0\\
1 & 0 & 0
\end{pmatrix},
\end{equation}

\begin{equation}
R_{\mathrm{sym},3} \equiv R_{[100]}\!\left(\frac{\pi}{2}\right) =
\begin{pmatrix}
1 & 0 & 0\\
0 & 0 & -1\\
0 & 1 & 0
\end{pmatrix},
\end{equation}

\begin{equation}
R_{\mathrm{sym},4} \equiv R_{[001]}\!\left(-\frac{\pi}{2}\right) = 
\begin{pmatrix}
0 & 1 & 0\\
-1 & 0 & 0\\
0 & 0 & 1
\end{pmatrix},
\end{equation}

\begin{equation}
R_{\mathrm{sym},5} \equiv R_{[010]}\!\left(-\frac{\pi}{2}\right) = 
\begin{pmatrix}
0 & 0 & -1\\
0 & -1 & 0\\
1 & 0 & 0
\end{pmatrix},
\end{equation}

\begin{equation}
R_{\mathrm{sym},6} \equiv R_{[100]}\!\left(-\frac{\pi}{2}\right) =  
\begin{pmatrix}
-1 & 0 & 0\\
0 & 0 & -1\\
0 & 1 & 0
\end{pmatrix}.
\end{equation}

It is convenient to express all intrinsic quantities of the ions in the laboratory basis. Given a matrix quantity $U$ relative to class 1, in order to find the corresponding matrix for class $i$, the symmetry operator is first expressed in the laboratory basis as
\begin{equation}
R'_{\mathrm{sym},i} = R_{\mathrm{lab}\rightarrow\mathrm{xtal}}^{\dagger} \, R_{\mathrm{sym},i} \, R_{\mathrm{lab}\rightarrow\mathrm{xtal}}.
\end{equation}
The transformed matrix is then obtained as
\begin{equation}\label{eq:matrix_trans}
U_i = R'_{\mathrm{sym},i} \, U \, {R'^{\dagger}_{\mathrm{sym},i}}.
\end{equation}
For vector quantities $\mathbf{v}$, the corresponding transformation is
\begin{equation}\label{eq:vector_trans}
\mathbf{v}_i = R'_{\mathrm{sym},i} \, \mathbf{v}.
\end{equation}

The matrix quantities transformed in this way are the $g$-tensors of the ground and excited states, ${g}_{\mathrm{g}}$ and ${g}_{\mathrm{e}}$, and the piezospectroscopic tensor $A$.  
The vectors that are transformed are the dipoles $(\mathbf{d}_{\parallel}(\mathbf{B}_0),\,\mathbf{d}_{\perp}(\mathbf{B}_0))$ and the reference magnetic field $\mathbf{B}_0$, which are intrinsic to the site.  
Quantities treated as external and therefore not transformed include the applied magnetic field $\mathbf{B}$, the electric field polarization $\boldsymbol{\epsilon}$, and the stress tensor $\sigma$.

\section{Determination of dipoles under an arbitrary magnetic field}\label{app_dipole_determination}

We describe how to determine the electric transition dipole elements
\((\mathbf{d}_{\parallel}(\mathbf{B}),\,\mathbf{d}_{\perp}(\mathbf{B}))\)
for an arbitrary magnetic field \(\mathbf{B}\), starting from known dipoles at field \(\mathbf{B}_0\).
The electric dipole operator $\hat{d}_k(\mathbf{B})$ is written in the spin basis 
\(\bigl\{\ket{\downarrow_{\mathrm{g}}},\ket{\uparrow_{\mathrm{g}}},\ket{\downarrow_{\mathrm{e}}},\ket{\uparrow_{\mathrm{e}}}\bigr\}_{\mathbf{B}}\),
defined by the magnetic field orientation \(\mathbf{B}(\varphi,\vartheta)\)~\cite{Raha2020}:
\begin{equation}
\hat{d}_k(\mathbf{B}) =
\begin{pmatrix}
0 & 0 & d_{k,\parallel}(\mathbf{B}) & d_{k,\perp}(\mathbf{B}) \\
0 & 0 & -d^{*}_{k,\perp}(\mathbf{B}) & d^{*}_{k,\parallel}(\mathbf{B}) \\
0 & 0 & 0 & 0 \\
0 & 0 & 0 & 0
\end{pmatrix}
+ \text{h.c.},
\end{equation}
where the index $k\in\{x,y,z\}$ labels the Cartesian components of the dipole operator
in the laboratory basis. Therefore, \(d_{k,\parallel}\) and \(d_{k,\perp}\) are the \(k\)-th components of
\(\mathbf{d}_{\parallel}\) and \(\mathbf{d}_{\perp}\), respectively. 
This matrix structure follows from the fact that the electric dipole operator is even under time reversal, which imposes relations between matrix elements within each Kramers doublet.
Magnetic dipole contributions are neglected, as their effect is estimated to be negligible in our system, consistent with the short measured bulk lifetime~\cite{Gritsch2022, Holzapfel2024}.

The states
\(\{\ket{\uparrow_\mathrm{g(e)}},\ket{\downarrow_\mathrm{g(e)}}\}_{\mathbf{B}}\)
diagonalize the spin Hamiltonians in Eq.~\eqref{eq:spin_hamiltonian}.
In this representation, the quantization axis corresponds to the direction of the effective magnetic field \(g_{\mathrm{g(e)}}\cdot\mathbf{B}\).
In the standard form of the Hamiltonian, written using the vector of Pauli matrices, the Hamiltonian is diagonal when the quantization axis points along \(z\).
We define the field satisfying this condition as \(\mathbf{B}_{\mathrm{ref}}\).
Diagonalization of \(\hat{H}_{g(e)}(\mathbf{B})\) yields the rotation matrices \(R_\mathrm{g(e)}(\mathbf{B})\) that transform from the reference spin basis (defined at \(\mathbf{B}_{\mathrm{ref}}\)) to the field-aligned basis.
The total \(4\times4\) transformation acting on the dipole operator is
\begin{equation}
R(\mathbf{B}) =
\bigl(P\, R_\mathrm{g}(\mathbf{B})\, P\bigr)
\oplus
\bigl(P\, R_\mathrm{e}(\mathbf{B})\, P\bigr),
\end{equation}
where the permutation matrix
\begin{equation}
P =
\begin{pmatrix}
0 & 1\\
1 & 0
\end{pmatrix}
\end{equation}
swaps the order of \(\ket{\downarrow_\mathrm{g(e)}}\) and \(\ket{\uparrow_\mathrm{g(e)}}\) states to maintain consistent indexing between ground and excited manifolds.
Starting from the known coupling matrix \(\hat{d}_k(\mathbf{B}_0)\), we first obtain its representation in the reference frame:
\begin{equation}
\hat{d}_k(\mathbf{B}_{\mathrm{ref}}) =
R(\mathbf{B}_0)\, \hat{d}_k(\mathbf{B}_0)\, R^{\dagger}(\mathbf{B}_0).
\end{equation}
For an arbitrary magnetic-field orientation, the dipole operator follows as
\begin{equation}\label{eq:dipole_from_ref}
\hat{d}_k(\mathbf{B}) =
R^{\dagger}(\mathbf{B})\, \hat{d}_k(\mathbf{B}_{\mathrm{ref}})\, R(\mathbf{B}),
\end{equation}
which determines \(\mathbf{d}_{\parallel}(\mathbf{B})\) and \(\mathbf{d}_{\perp}(\mathbf{B})\).

\section{Strain-orbital coupling model}\label{app_strain_orbital_coupling}
The coupling between local lattice deformation and optical transition is
described by the piezospectroscopic tensor $A$. We note that the underlying interaction is with the local strain field, but in the linear elastic regime the strain tensor $\varepsilon$ and the stress tensor $\sigma$ are related by the Si elasticity tensor.  
The coupling can therefore equivalently be written in terms of the stress
tensor, and the symmetry-allowed form of the coupling matrix is identical in
both representations.
Following the standard piezospectroscopy convention, we parametrize the
frequency shift in terms of the stress tensor, so that the coefficients $A_{ij}$
are expressed in units of Hz/Pa, as in Refs.~\cite{Kaplyanskii1967, Buzzi2025, Louchet2019}.  

Eq.~(\ref{eq:strain_coupling}) can be simplified following the
piezospectroscopic model for non-cubic centers in cubic crystals~\cite{Kaplyanskii1967}.
For a $C_{2v}$ point group in an $O_h$ host, the piezospectroscopic tensor takes
the form
\begin{equation}
A =
\begin{pmatrix}
A_2 & A_3 & 0\\
A_3 & A_2 & 0\\
0 & 0 & A_1
\end{pmatrix}.
\end{equation}
The strain-induced energy shift is then given by
\begin{equation}\label{eq:simpl_strain}
\Delta_{\mathrm{strain}} =
A_1\sigma_{zz} + A_2(\sigma_{xx} + \sigma_{yy}) + 2A_3\sigma_{xy},
\end{equation}
where both $\sigma$ and $A$ are expressed in the crystal frame.
Eq.~\eqref{eq:simpl_strain} applies only to class~1 ions, which are aligned with the crystal axes.  
For other magnetic classes, the tensor must be transformed using the corresponding symmetry operation $R_{\mathrm{sym},i}$:
\begin{equation}
A_i = R_{\mathrm{sym},i}\, A\, R_{\mathrm{sym},i}^{\dagger}.
\end{equation}

The strain tensor $\varepsilon$ is obtained from finite-element simulations of suspended waveguides oriented along direction~$\alpha$, and the stress tensor $\sigma$ is computed from $\varepsilon$ using the Si elasticity tensor~\cite{SM}.
From the fit of the ensemble measurements (Fig.~\ref{fig2}c), we obtain the coefficients  
$A_{1} = 3.34 \pm 0.48$~Hz/Pa,  
$A_{2} = -0.50 \pm 0.50$~Hz/Pa, and  
$A_{3} = -7.30 \pm 0.21$~Hz/Pa.  
After diagonalizing $A$ with $R_{\mathrm{lab}\rightarrow\mathrm{xtal}}$, we find it expressed in the laboratory basis:
\begin{equation}
A_{\mathrm{lab}}=
\begin{pmatrix}
-7.8(3) & 0 & 0\\
0 & 6.8(3) & 0\\
0 & 0 & 3.3(5)
\end{pmatrix}
\ \mathrm{Hz/Pa}.
\end{equation}

\section{Experimental details}\label{app_exp_parameters}
\begin{figure}[htb!]
\centering
\includegraphics[]{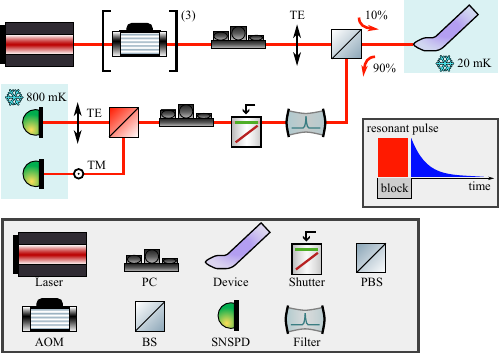}
\caption{\textbf{Simplified optical setup and time sequence.}
A tunable CW laser is pulsed by three acousto–optical modulators (AOMs), passes through a 90/10 beam splitter (BS), and is sent to the device with TE-mode polarization. The reflected signal passes through a 0.5~nm broad filter at the ZPL and a polarizing beam splitter (PBS) that separates the TE and TM components, which are then directed to superconducting nanowire single-photon detectors (SNSPDs). Polarization controllers (PCs) are used to align the polarization. A shutter blocks the strong reflected excitation pulse to protect the SNSPDs. The setup for ensemble measurements employs two AOMs as a shutter, while in cavity-enhanced measurements a high-speed optical switch is used. In addition, for cavity-enhanced measurements, the 90/10 BS is replaced by a 99/1 BS to increase efficiency, and the PBS is replaced by a 50/50 BS to perform HBT measurements. No filter is used in the cavity-enhanced measurements. The full setups are reported in~\cite{SM}.}
\label{ext_fig_setup}
\end{figure}

A simplified schematic of the setup for ensemble measurements is shown in Fig.~\ref{ext_fig_setup}, while a complete description is provided in~\cite{SM}. 

A resonant optical pulse excites the transition, followed by time-resolved detection of the PL signal after polarization (TE or TM) and spectral filtering ($\sim$0.5~nm) to isolate the ZPL. 
The key experimental parameters are the pulse duration $t_{\mathrm{pulse}}$, the cycle period $t_{\mathrm{cycle}}$, the total integration time per frequency point $t_{\mathrm{tot}}$, the PL integration window $[t_{\mathrm{min}}, t_{\mathrm{max}}]$, and the instantaneous pulse power $P$.

For the ensemble measurements in LWGs, the parameters are $t_{\mathrm{pulse}} = 0.1$~ms, $t_{\mathrm{cycle}} = 1.5$~ms, $t_{\mathrm{tot}} = 3$~s, $t_{\mathrm{min}} = 0.104$~ms, and $t_{\mathrm{max}} = 0.6$~ms.  
The excitation power is $P = \SI{4.4}{\micro\watt}$, well above the saturation power~\cite{SM}.

For the cavity-enhanced measurements in NBs, we use $t_{\mathrm{pulse}} = \SI{10}{\micro\second}$, $t_{\mathrm{cycle}} = \SI{50}{\micro\second}$, $t_{\mathrm{min}} = \SI{11.6}{\micro\second}$, and $t_{\mathrm{max}} = \SI{16}{\micro\second}$.  
The total integration time is $t_{\mathrm{tot}} = 10$~min for the measurements in Fig.~\ref{fig4}a, Fig.~\ref{fig5}b and e; 10~s for those in Fig.~\ref{fig4}d and e, and Fig.~\ref{fig5}a and d; and 1~h for the Hanbury Brown and Twiss (HBT) measurement in Fig.~\ref{fig4}c.  
The excitation power is $P = 1$~nW in Fig.~\ref{fig4}a and $P = 0.4$~nW in the other measurements, both below the saturation power of 1.3~nW~\cite{SM}.

We highlight the main differences between the two HBT measurements in Fig.~\ref{fig4}a and c. 
The measurement in Fig.~\ref{fig4}a was acquired over 10~minutes with an excitation power of 1~nW, with the laser frequency stabilized to a wavelength meter. 
In contrast, the measurement in Fig.~\ref{fig4}c was integrated for 1~hour with an excitation power of 0.4~nW, and the laser was locked to a reference cavity, providing a more accurate characterization of the second-order correlation function.

\section{Model for polarization-resolved PL intensity}
We excite the ensemble of ions using $y$-polarized light coupled to the TE mode of the waveguide. Since the excitation power is well above saturation, the input polarization does not influence the excitation probability~\cite{SM}. However, the optical response still depends on whether the laser is resonant with the spin-preserving or spin-flip transition: due to finite cyclicity, the addressed ground state can be depleted over time through decay into the opposite spin state.
We define 
\begin{equation}\label{eq:coupling_specific}
    F_{\mathrm{TE},\parallel} = |\mathbf{d}_\parallel \cdot \boldsymbol{\epsilon}_{\mathrm{TE}}|^2,
\end{equation}
where $\boldsymbol{\epsilon}_{\mathrm{TE}}$ is the electric-field polarization for the TE mode.  
We define $F_{\mathrm{TM},\parallel}$, $F_{\mathrm{TE},\perp}$, and $F_{\mathrm{TM},\perp}$ analogously.
The excitation probability is proportional to $F_{\mathrm{TE},\parallel}+F_{\mathrm{TM},\parallel}$ for spin-preserving transitions, and to $F_{\mathrm{TE},\perp}+F_{\mathrm{TM},\perp}$ for spin-flip transitions (Eq.~\eqref{eq:Fermi_golden}).
The collected PL includes both types of transitions, since they fall within the spectral filter linewidth.
The TE and TM components are separated at the output. When TE light is selected, the detected signal is proportional to $F_{\mathrm{TE},\parallel}+F_{\mathrm{TE},\perp}$.  
For TM detection, because the mirror at the waveguide end reflects only TE modes, only photons propagating toward the taper are collected, introducing an additional factor of $1/2$.
Combining these considerations, the measured intensities for the four detection configurations are:
\begin{equation}\label{eq:PL_direct_link}
\begin{aligned}
I_{\mathrm{TE},\parallel} &\propto (F_{\mathrm{TE},\parallel}+F_{\mathrm{TM},\parallel}) (F_{\mathrm{TE},\parallel}+F_{\mathrm{TE},\perp}), \\
I_{\mathrm{TE},\perp} &\propto (F_{\mathrm{TE},\perp}+F_{\mathrm{TM},\perp}) (F_{\mathrm{TE},\parallel}+F_{\mathrm{TE},\perp}), \\
I_{\mathrm{TM},\parallel} & \propto \tfrac{1}{2}(F_{\mathrm{TE},\parallel}+F_{\mathrm{TM},\parallel}) (F_{\mathrm{TM},\parallel}+F_{\mathrm{TM},\perp}), \\
I_{\mathrm{TM},\perp} &\propto \tfrac{1}{2} (F_{\mathrm{TE},\perp}+F_{\mathrm{TM},\perp}) (F_{\mathrm{TM},\parallel}+F_{\mathrm{TM},\perp}).
\end{aligned}
\end{equation}
These expressions directly relate the measured PL intensities to the dipole moments and the optical modes.

\section{Procedure for computing magnetic-field-dependent PL colormaps}
\label{app_workflow}

\begin{figure}[htb!]
\centering
\includegraphics[]{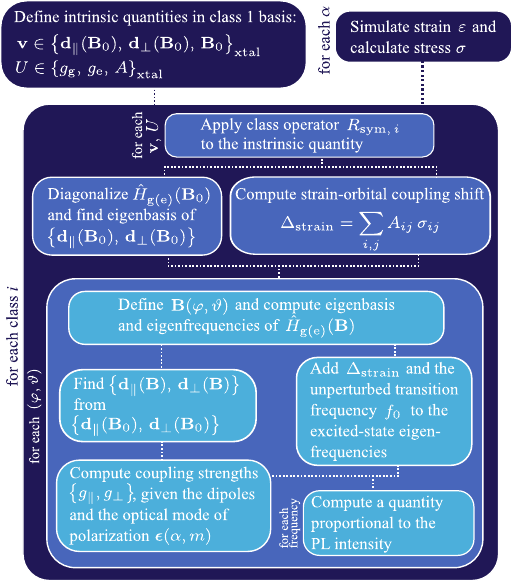}
\caption{\textbf{Flow diagram of the theoretical model.}
For each magnetic class $i$, the model computes a quantity proportional to the PL intensity as a function of excitation frequency and magnetic-field orientation $(\varphi, \vartheta)$. Starting from the spin Hamiltonians and strain-induced shifts, the transition frequencies and dipoles are evaluated, from which the coupling strengths to the optical modes are obtained. The PL spectrum is then constructed as a sum of Lorentzian lines centered at the transition frequencies with amplitudes given by Eq.~\eqref{eq:PL_direct_link}. For ensemble measurements, the total PL is obtained by summing the contributions from all classes with equal weight.
}
\label{ext_fig_model}
\end{figure}

\begin{figure*}[htb!]
\centering
\includegraphics[]{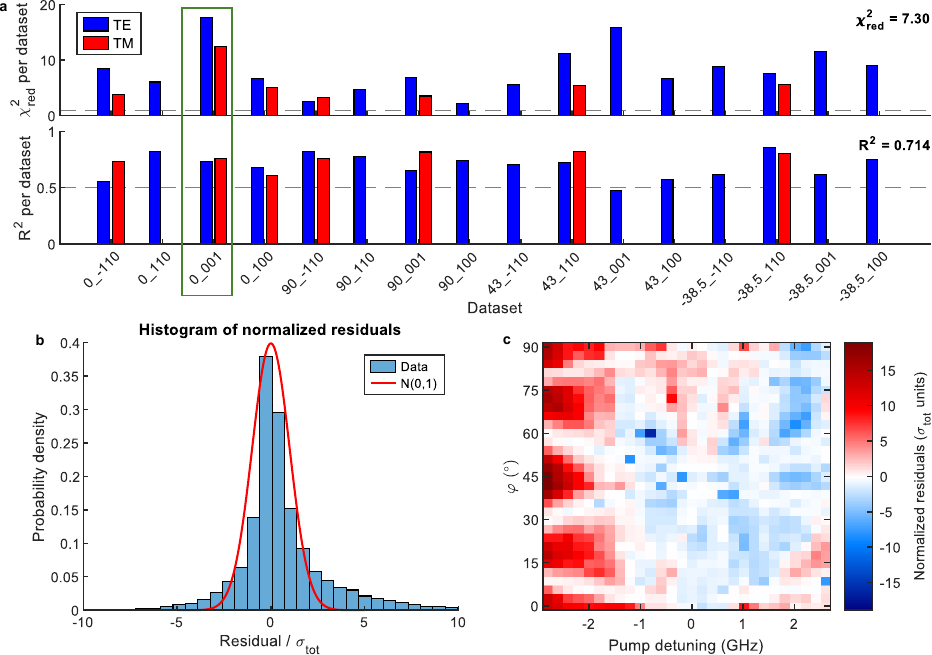}
\caption{\textbf{Ensemble fit metrics.}
\textbf{a.} Reduced $\chi^2$ and coefficient of determination $R^2$ for each dataset, as defined in Appendix~\ref{app_fit_goodness}. The x-axis labels indicate the waveguide orientation $\alpha$ (in degrees) and the plane of magnetic field rotation. Blue and red bars correspond to TE and TM modes, respectively. In the top panel, the dashed line marks $\chi^2_{\mathrm{red}} = 1$; in the bottom panel, the dashed line marks $R^2 = 0.5$. Global $\chi^2_{\mathrm{red}}$ and $R^2$ are indicated.
\textbf{b.} Histogram of normalized residuals with a standard normal distribution (mean 0, variance 1) overlaid. 
\textbf{c.} Normalized residuals for the TE mode dataset highlighted by the green rectangle in \textbf{a}.}
\label{ext_fig_stats}
\end{figure*}

In Fig.~\ref{ext_fig_model}, we outline the procedure used to construct the model colormaps shown in Fig.~\ref{fig2}c (right panels). Specifically, we compute a quantity proportional to the PL intensity (Eq.~\eqref{eq:PL_direct_link}) as a function of excitation frequency and magnetic-field orientation $(\varphi, \vartheta)$.
We consider a given magnetic class $i$. Intrinsic quantities are defined in the crystal frame and transformed to the corresponding class frame using the symmetry operator $R_{\mathrm{sym},i}$ (Eqs.~\eqref{eq:matrix_trans}, \eqref{eq:vector_trans}).
For each magnetic-field orientation $\mathbf{B}(\varphi,\vartheta)$, the ground- and excited-state spin Hamiltonians (Eq.~\eqref{eq:spin_hamiltonian}) are diagonalized to obtain eigenstates and eigenfrequencies. For a given waveguide orientation $\alpha$, the average strain $\varepsilon$ is simulated and the corresponding stress $\sigma$ is inferred. The resulting strain-induced energy shifts $\Delta_{\mathrm{strain}}$, computed using the piezospectroscopic tensor $A$ (Eq.~\eqref{eq:strain_coupling}), are added to the unperturbed transition frequency $f_0$ and to the eigenfrequencies of the excited-state spin Hamiltonian.
The field-dependent transition dipoles are evaluated from the eigenstates and from the reference dipoles $(\mathbf{d}_{\parallel}(\mathbf{B}_0),\,\mathbf{d}_{\perp}(\mathbf{B}_0))$ using Eq.~\eqref{eq:dipole_from_ref}. The spin-preserving and spin-flip coupling strengths $(g_{\parallel}, g_{\perp})$ are then obtained from their projections onto the optical mode polarization $\boldsymbol{\epsilon}(\alpha,m)$ (Eqs.~\eqref{eq:Fermi_golden},~\eqref{eq:coupling_specific}).
The expected PL spectrum for class $i$ is constructed as a sum of Lorentzian lines centered at the four transition frequencies, with amplitudes given by the corresponding expressions in Eq.~\eqref{eq:PL_direct_link}, as determined from the coupling strengths.
Finally, the model PL colormaps shown in Fig.~\ref{fig2}c (right panels) are obtained by summing the contributions from all six magnetic classes with equal weight.

\section{Fitting procedure and metrics}\label{app_fit_goodness}

\begin{figure*}[htb!]
\centering
\includegraphics[]{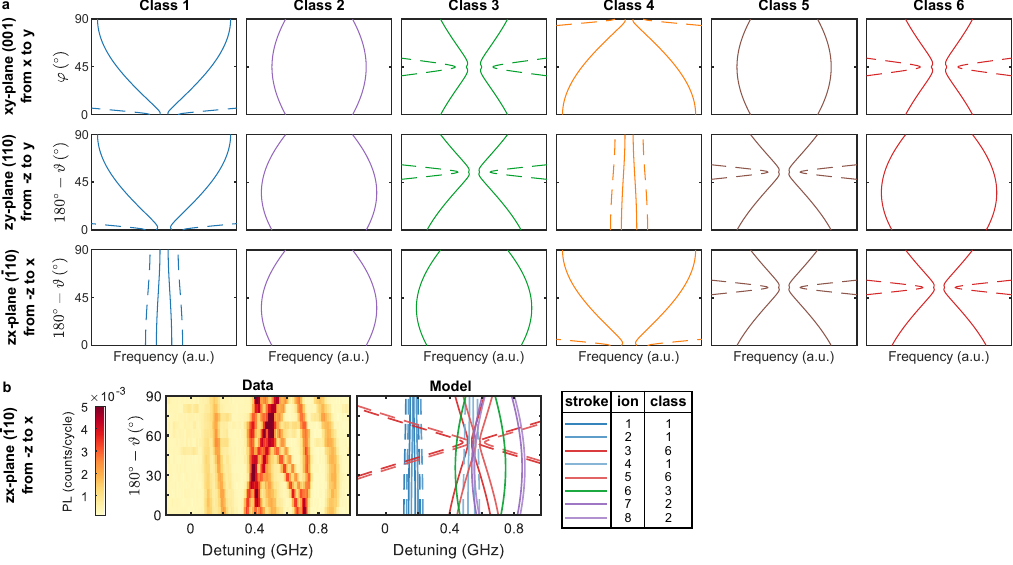}
\caption{\textbf{Individual magnetic-class transition frequencies and comparison with experiment.}
\textbf{a.} Calculated field-dependent transition frequencies for each magnetic class. Solid (dashed) curves indicate spin-preserving (spin-flip) transitions. In the first row, the magnetic field is rotated within the (001) plane ($\vartheta=\pi/2$) with $\varphi$ varying from $0$ to $\pi/2$. In the second row, the field is rotated in the (110) plane ($\varphi=\pi/2$) with $\vartheta$ varying from $\pi$ to $\pi/2$. In the third row, the field is rotated in the $(\bar{1}10)$ plane ($\varphi=0$) with $\vartheta$ varying from $\pi$ to $\pi/2$. Any two of these rotation planes are sufficient to uniquely identify the magnetic class.
\textbf{b.} Comparison between experiment and model for field rotations in the $(\bar{1}10)$ plane, analogous to Fig.~\ref{fig4}d,e. The experimental data are consistent with the model derived from the other two rotation planes.
}
\label{ext_fig_classes}
\end{figure*}

In the procedure described in Appendix~\ref{app_workflow}, the only unknown quantities are $\mathbf d_{\parallel}(\mathbf B_0)$, $\mathbf d_{\perp}(\mathbf B_0)$, and $A$.
These parameters are determined by fitting the modeled PL intensity to the measured data across all datasets, by minimizing the chi-square function
\begin{equation}
\chi^2 = \sum_i \frac{\bigl(I_{\mathrm{data},\,i} - I_{\mathrm{mod},\,i}\bigr)^2}{\sigma_i^2},
\end{equation}
where $I_{\mathrm{data},\,i}$ and $I_{\mathrm{mod},\,i}$ are the normalized measured and modeled PL intensities for data point $i$, and $\sigma_i$ are the corresponding uncertainties.
Normalization is performed independently for each dataset by subtracting the background (defined as the minimum signal) and dividing by the maximum signal.

Another figure of merit is the coefficient of determination $R^2$, defined as
\begin{equation}
R^2 = 1 - \frac{\chi^2}{\chi^2_{\text{const}}},
\end{equation}
with
\begin{equation}
\chi^2_{\text{const}} = \sum_i \frac{(I_{\mathrm{data},\,i} - \bar{I}_{\mathrm{data}})^2}{\sigma_i^2},
\end{equation}
where $\bar{I}_{\mathrm{data}}$ is the weighted mean of the data. 
This quantity measures the fraction of the variance explained by the model relative to a constant (mean) model, and thus quantifies how well the model captures the structure of the PL signal beyond a uniform background.

The error bars used in the fit include both the uncertainties in the PL intensity ($\sigma_{I}$) and in the angle ($\sigma_{\vartheta}$, $\sigma_{\varphi}$). The PL uncertainty follows Poissonian statistics. The angular uncertainty is more difficult to estimate, as it depends on several factors: the current source stability, the magnetic field calibration, and possible stray magnetic fields~\cite{SM}. The angular resolution in our measurements is approximately $3^\circ$, which we therefore assign as the uncertainty on each data point. The total propagated uncertainty is calculated as
\begin{equation}
\sigma_{\text{tot}}^{2} = \sigma_{I}^{2} +
\left( \frac{\partial I}{\partial \vartheta} \sigma_{\vartheta} \right)^{2} +
\left( \frac{\partial I}{\partial \varphi} \sigma_{\varphi} \right)^{2},
\end{equation}
where $I = I_{\mathrm{mod},\, i}$.
Since analytic expressions for $\partial I / \partial \vartheta$ and $\partial I / \partial \varphi$ are not available, these derivatives are computed numerically as
\begin{equation}
\frac{\partial I}{\partial \vartheta} \approx
\frac{I(\vartheta + \delta) - I(\vartheta - \delta)}{2\delta},
\end{equation}
using a small step $\delta \approx 10^{-3}$~rad.

Additional details on the fitting procedure, as well as the best-fit parameters and their uncertainties, are reported in~\cite{SM}. The global fit yields a reduced $\chi^2$ of 7.3, indicating deviations beyond purely statistical noise, and a coefficient of determination $R^2 = 0.71$, showing that the model captures a significant fraction of the observed variance.
The figures of merit for each dataset are shown in Fig.~\ref{ext_fig_stats}a. In some cases, the values deviate significantly from the ideal model. Analysis of the residuals (Fig.~\ref{ext_fig_stats}c) indicates that most of the discrepancy arises from an uneven background, likely due to PL from ions belonging to another site~\cite{SM}. The tails observed in the histogram of normalized residuals (Fig.~\ref{ext_fig_stats}b) further suggest the presence of a small systematic error.

\section{Magnetic class identification}\label{app_class_identification}

In Fig.~\ref{ext_fig_classes}a, we show the calculated field-dependent transition frequencies for each magnetic class. Each row corresponds to magnetic-field rotations in one of the $(001)$, $(110)$, or $(\bar{1}10)$ planes. Within a given plane, two pairs of classes produce identical spectra and cannot be distinguished. However, the indistinguishable pairs differ between planes, so combining measurements in any two planes is sufficient to uniquely identify the magnetic class.

In Fig.~\ref{ext_fig_classes}b (left), we report, for completeness with Fig.~\ref{fig4}d,e, the measured PL colormaps of the cavity-enhanced ions for magnetic-field rotation in the $(\bar{1}10)$ plane. The theoretical spectral lines in Fig.~\ref{ext_fig_classes}b (right), obtained from the other two planes, are in good agreement with the data.

We now describe the procedure used to identify the ions and assign their magnetic class. In the measured colormaps (e.g., Fig.~\ref{ext_fig_classes}b, left), we identify sets of two or four spectral lines—depending on whether spin-flip transitions are visible—that split around a common central frequency. This allows us to identify eight ions in our dataset. For each ion, the observed spectral lines are then compared with the six theoretical patterns shown in Fig.~\ref{ext_fig_classes}a for at least two rotation planes. This procedure enables an unambiguous assignment of the magnetic class to each ion, from which the predicted spectral lines in Fig.~\ref{ext_fig_classes}b (right) are obtained.

\section{Lifetime data processing}\label{app_lifetime_data_processing}
For the characterization of the lifetime as a function of the magnetic field (Fig.~\ref{fig5}), we optimize the total acquisition time by recording data only at the frequencies corresponding to the spectral peaks. However, the exact peak positions cannot be known a priori because of laser locking accuracy, magnetic field uncertainty, and spectral diffusion. For each magnetic field setting, we first perform a fast frequency scan to locate the PL peaks (Fig~\ref{fig5}a,d). Once identified, we repeat PL measurements at three frequencies for each peak: the central frequency and two points detuned by $\pm10$~MHz, with an integration time of 10~min per point.
We measure two branches corresponding to the two spin-preserving transitions. For each branch, the data point with the highest PL among the three frequency detunings is selected. Its time-resolved PL trace, $I(t)$, is then fitted to an exponential decay function:
\begin{equation}
    I(t) = I_0\, e^{-(t-t_0)/\tau_0} + N,
\end{equation}
where $I_0$ is the signal amplitude at $t=t_0$, $t_0=\SI{10}{\micro\second}$ marks the end of the excitation pulse, $\tau_0$ is the lifetime, and $N$ is the background noise level. The lifetime is related to the emission rate in Eq.~(\ref{eq:enhanced_emission_B_field}) by $\tau_0 = 1/ \,\Gamma_\mathrm{P}$.

All measurements are performed using two detectors in a HBT configuration, which allows simultaneous extraction of the second-order correlation function $g^{(2)}(\tau)$.

\clearpage
\onecolumngrid

% Supplemental Material appended for arXiv submission
\setcounter{section}{0}
\setcounter{subsection}{0}
\setcounter{figure}{0}
\setcounter{table}{0}
\setcounter{equation}{0}

\renewcommand{\thesection}{S\arabic{section}}
\renewcommand{\thesubsection}{\thesection.\arabic{subsection}}
\renewcommand{\thefigure}{S\arabic{figure}}
\renewcommand{\thetable}{S\arabic{table}}
\renewcommand{\theequation}{S\arabic{equation}}

% IMPORTANT: unique hyperref anchors
\renewcommand{\theHsection}{S.\arabic{section}}
\renewcommand{\theHsubsection}{S.\arabic{section}.\arabic{subsection}}
\renewcommand{\theHfigure}{S.\arabic{figure}}
\renewcommand{\theHtable}{S.\arabic{table}}
\renewcommand{\theHequation}{S.\arabic{equation}}

\makeatletter
\setcounter{NAT@ctr}{0}
\makeatother

\begin{center}
{\Large \textbf{Supplemental Material}}\\[0.5em]
{\large \textbf{A unified framework for determining transition dipole polarization\\ in solid-state spin defects}}
\end{center}
\vspace{1em}

\section{Device details and experimental setup}\label{sec:setup}

    \subsection{Long waveguide design}
    
\begin{figure*}[htb!]
    \centering
    \includegraphics[]{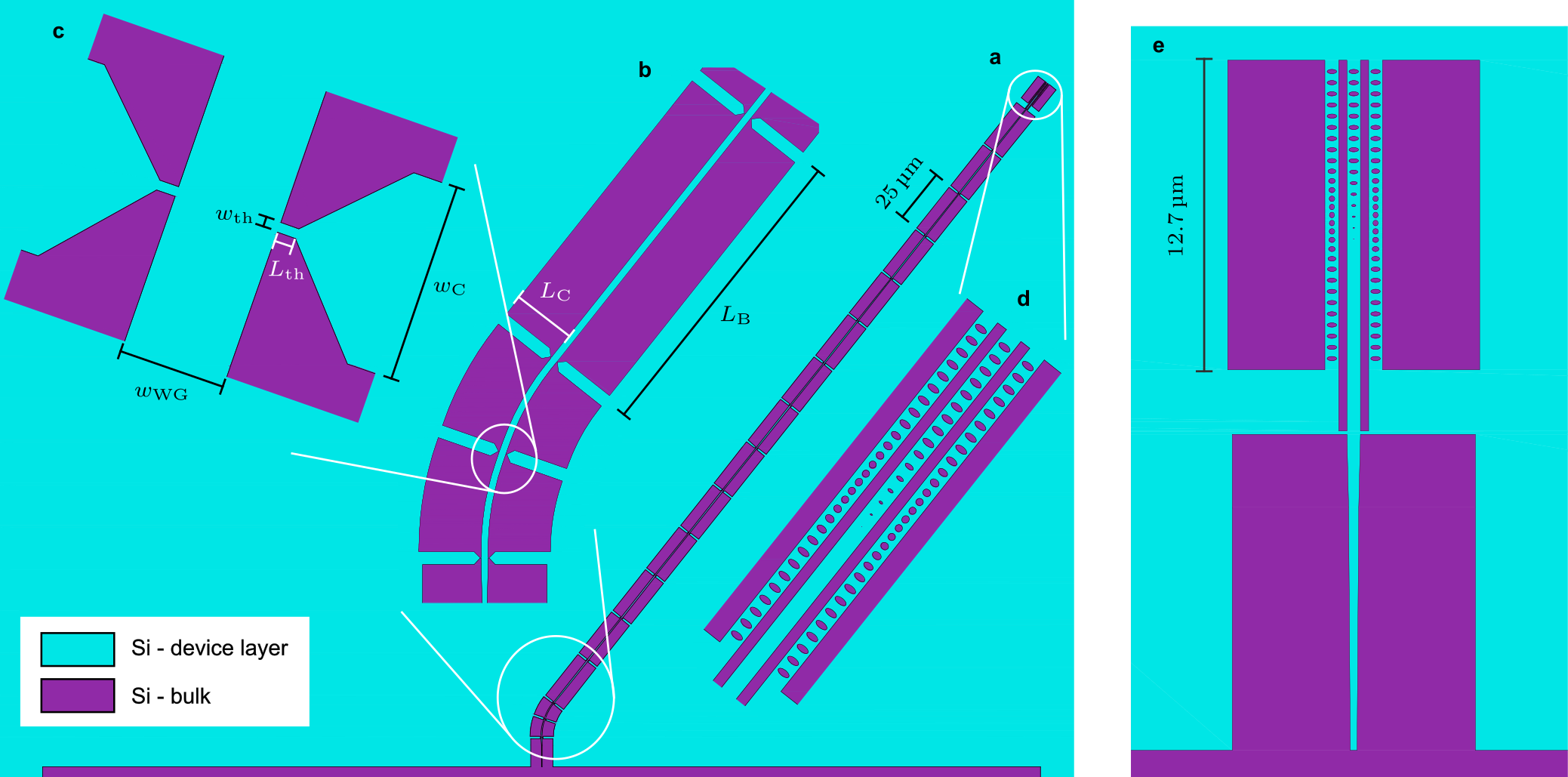}
    \caption{\textbf{Long waveguide and nanobeam designs.}
    \textbf{a.} Design of a long waveguide (LWG) with $\alpha = -38.5^\circ$. 
    \textbf{b.} Zoom of the bending region. 
    \textbf{c.} Zoom of the tether geometry. 
    \textbf{d.} Zoom of the waveguide end, including one photonic crystal (PhC) mirror and two evanescently coupled PhC cavities. 
    \textbf{e.} Design of two nanobeams (NBs) evanescently coupled to a waveguide, consisting of a coupling taper followed by a PhC mirror, similarly to \textbf{d}.
    }
    \label{figSI:device_design}
    \end{figure*}

    The design of a long waveguide (LWG) with orientation angle $\alpha = -38.5^\circ$ is shown in Fig.~\ref{figSI:device_design}a-d. 
    
    The suspended waveguide requires many tethers for mechanical support, but their number must be minimized to reduce optical losses. 
    Likewise, the tether width should be made as small as possible to limit scattering, while remaining mechanically robust.

    Because the chip is diced prior to oxide release, all coupling tapers must be aligned in parallel so that each device can be accessed with a lensed fiber inside the cryostat. 
    For this reason, we cannot simply fabricate straight waveguides rotated by an angle. 
    Instead, the waveguide is straight near the taper, and the bend is introduced immediately after the taper so that its contribution is negligible compared to the long segment oriented at the desired angle.
    
    Particular care is required in the bending region, where an extra tether is included to improve mechanical stability. 
    We choose a bending radius much larger than the optical wavelength in silicon, ensuring negligible bending loss. 
    This is confirmed experimentally: all LWGs, regardless of their bending angle, exhibit coupling efficiencies between $32\%$ and $38\%$, where the efficiency is defined as $\sqrt{P_{\mathrm{out}}/P_{\mathrm{in}}}$, with $P_{\mathrm{in}}$ the TE input power at the coupling taper and $P_{\mathrm{out}}$ the reflected power measured after a circulator. 
    Device-to-device variations mainly arise from edge irregularities due to dicing and from fabrication tolerances.
    
    The suspended structures are released using vapor hydrofluoric acid, so the surrounding environment never undergoes a liquid–gas phase transition.  
    This prevents stiction and collapse of the long suspended waveguides.
    
    Finally, each LWG incorporates two evanescently coupled photonic-crystal (PhC) cavities—analogous to nanobeams (NBs)—next to the PhC mirror.
    These cavities are intentionally detuned from the Er:Si optical transitions and serve only to unambiguously identify each device during characterization.

    \begin{table}[htb!]
        \caption{\textbf{Long waveguide design parameters.} 
        The geometric parameters follow the definitions in Fig.~\ref{figSI:device_design}. 
        Waveguides with $\alpha < 40^\circ$ are designed with bending radius $r_{\mathrm{b1}}$, whereas those with $\alpha > 40^\circ$ are designed with $r_{\mathrm{b2}}$. 
        The bending radius is expressed in units of the effective wavelength in silicon, where $\lambda = 1550$~nm and the refractive index is $n = 3.48$.
        }
        \label{tabSI:LWG_parameters}
        \setlength{\tabcolsep}{10pt} % default is ~6pt — increase to add spacing
        \begin{tabular}{lccccccccc}
        \toprule
        $N_\mathrm{B}$ & $L_\mathrm{B}$ (\SI{}{\micro\meter}) & $w_{\mathrm{WG}}$ (\SI{}{\micro\meter}) &
        $w_{\mathrm{th}}$ (nm) & $L_{\mathrm{th}}$ (\SI{}{\micro\meter}) &
        $w_\mathrm{C}$ (\SI{}{\micro\meter}) & $L_\mathrm{C}$ (\SI{}{\micro\meter}) &
        $r_\mathrm{b1}$ ($\lambda/n$) & $r_\mathrm{b2}$ ($\lambda/n$) \\
        \midrule
        $14$ & $25$ & $0.55$ & $55$ & $0.1$ & $1$ & $5$ & $50$ & $10$ \\
        \botrule
        \end{tabular}
    \end{table}
    
    % \subsection{Long waveguides polarization maintenance}
    % \remark{Tests with optimized reflection. Moved to next device, polarization is maintained.}

    \subsection{Nanobeam design}
    
    Fig.~\ref{figSI:device_design}e shows a device consisting of a waveguide evanescently coupled to two PhC NBs, allowing two cavities to be integrated within a single waveguide.
    Fabrication imperfections typically shift the resonance wavelengths of the two cavities slightly relative to one another. 
    The NBs are designed and optimized using finite-element simulations following the approach of Ref.~\cite{Chan2012}. 
    For our design, the simulated cavity quality factor is $7.7 \times 10^6$.

    \subsection{Cryogenic setup}
\begin{figure*}[htb!]
    \centering
    \includegraphics[]{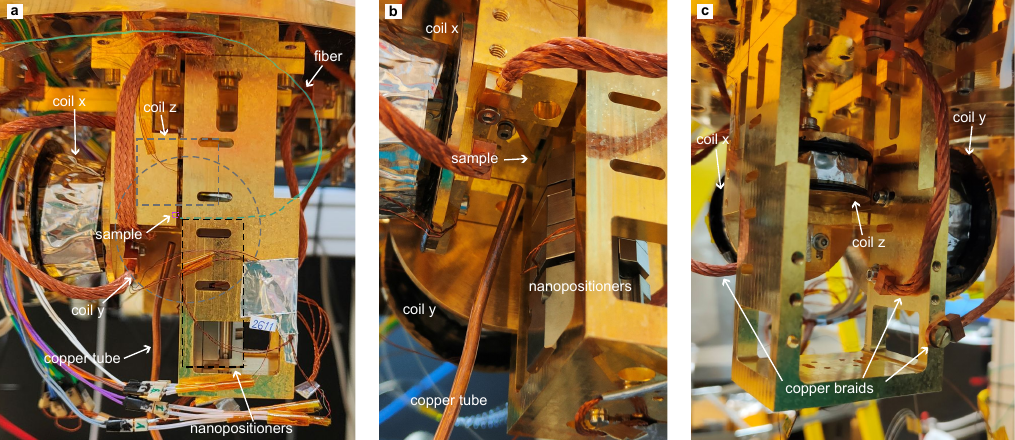}
    \caption{\textbf{Setup at the mixing chamber of the dilution refrigerator.} 
    The setup includes the sample, a stack of nanopositioners used to align the lensed fiber, a three-axis vector magnet, and a copper tube for gas tuning. 
    \textbf{a.} Side view of the setup inside the cryostat. 
    The lensed fiber (not visible) faces the sample (also not visible), located at the intersection of the vector magnet. The copper tube is oriented towards the sample. 
    \textbf{b.} View of the sample mounted upside down, together with the nanopositioner stack. 
    The lensed fiber is mounted on top of the stack and extends less than 1~mm beyond its edge, making it barely visible.
    \textbf{c.} Rear view of the three-axis vector magnet before installation of the sample and nanopositioners. 
    Copper braids are attached to improve thermal anchoring.}
    \label{figSI:fridge}
    \end{figure*}

    As shown in Fig.~\ref{figSI:fridge}, the sample is mounted on a gold-plated copper holder anchored to the mixing-chamber plate of a dilution refrigerator (Bluefors). 
    The holder incorporates a homemade three-axis vector magnet~\cite{DaPrato2025}, with thermal anchoring facilitated by copper braids. 
    The sample is mounted upside down to leave access to a copper tube, thermally anchored at the 4~K stage, which is used for gas tuning (see Sec.~\ref{sec:gas_tuning}). 
    
    A lensed fiber is placed on top of a stack of three nanopositioners (Attocube), which allow precise alignment in all spatial directions to optimize the coupling of light into the selected device. 
    For the ensemble measurements, a slightly modified setup is used, as described in Ref.~\cite{DaPrato2025}, where the copper tube is absent.

    \subsection{Laser light preparation}
\begin{figure*}[htb!]
    \centering
    \includegraphics[]{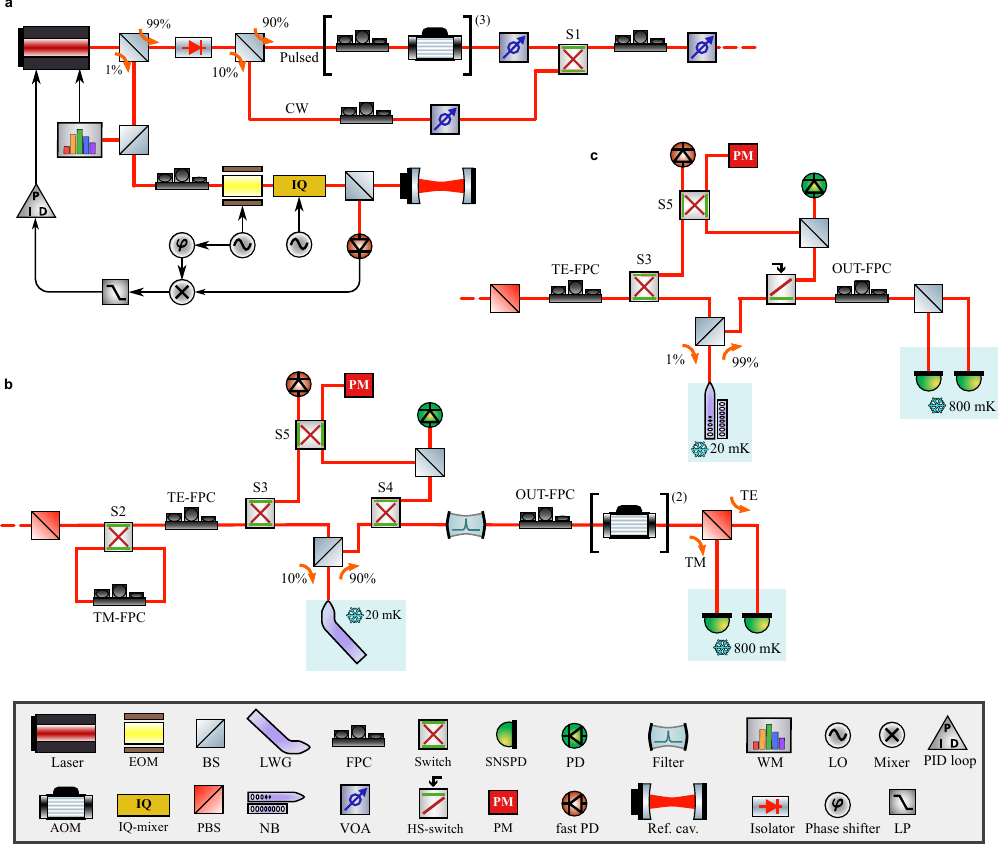}
    \caption{\textbf{Comprehensive optical setup.}
    Legend: AOM = acousto-optic modulator, EOM = electro-optic modulator, BS = beam splitter, PBS = polarizing beam splitter, LWG = long waveguide, NB = nanobeam, HS-switch = high-speed optical switch, SNSPD = superconducting nanowire single-photon detector, PM = optical power meter, PD = photodiode, Ref.~cav.~= reference cavity, WM = wavelength meter, LO = local oscillator, LP = low-pass filter, PID = proportional–integral–derivative controller.
    \textbf{a} Laser-source and preparation section, where the light is generated, frequency-stabilized, power-controlled, and optionally pulsed before being routed to the experiment. 
    This block is common to all measurements, with its continuation shown in \textbf{b} for the ensemble measurements in LWGs and in \textbf{c} for the cavity-enhanced measurements in NBs.
    \textbf{b}–\textbf{c} Input and output polarizations are aligned by means of TE-FPC, TM-FPC, and OUT-FPC. 
    Detectors (PDs and fast PDs) and a PM are used to optimize the time sequence alignment and optical coupling before performing the fluorescence measurements.
    } 
    \label{figSI:setup}
    \end{figure*}

    In Fig.~\ref{figSI:setup} we report the full optical setup used in the experiments. 
    Panel~a shows the section common to all measurements and includes the laser source preparation; 
    panel~b shows its continuation for the ensemble measurements in LWGs; 
    panel~c shows the continuation for the cavity-enhanced measurements in NBs.
    
    A continuous-wave (CW) laser (Toptica CTL) is frequency stabilized either to a wavelength meter (WM, HighFinesse) for most measurements or to a reference cavity (Stable Laser Systems) for the magnetic-field-dependent lifetime measurements.
    For cavity locking, the laser is first coarsely stabilized to the WM and subsequently locked to the cavity using the Pound–Drever–Hall technique.
    An optical IQ mixer enables frequency offsets from the cavity resonance, allowing the laser to be stabilized at arbitrary frequencies, with the absolute frequency accuracy ultimately limited by the WM.
    Three acousto-optic modulators (AOMs, G\&H) are used to pulse the laser with a sufficiently high extinction ratio, and variable optical attenuators (VOAs) control the optical power. 
    The optical switch S1 allows the AOMs to be bypassed, enabling CW operation of the laser, which is used for device characterization.
    
    \subsection{Cavity reflection measurements}
    
    Device characterization is performed via reflection measurements, used both to optimize the coupling between the lensed fiber and the device and to identify the device by its resonance frequency.  
    These measurements use the CW path shown in panel~a.  
    Light reflected from the device (panels~b and c) is routed to a photodetector (PD) via switch S4.  
    By scanning the laser wavelength, we obtain the cavity resonance profile (see also Sec.~\ref{sec:gas_tuning}).
    
    \subsection{Polarization alignment}
    
    Polarization control is crucial for the ensemble measurements in LWGs.  
    As shown in Fig.~\ref{figSI:setup}b, two optical paths can be selected using the switch S2.  
    We first select the path that bypasses the TM fiber polarization controller (TM-FPC).  
    A power meter (PM) measures the reflected power, and TE-FPC is adjusted to maximize this signal.  
    Since the Bragg mirror of the LWG is optimized for TE mode, this procedure ensures TE excitation at the input.
    
    We then select the path that passes through TM-FPC and minimize the reflected power at the PM, thereby selecting TM excitation.  
    Thus, by toggling S2, we can automatically switch between TE and TM input polarization.
    
    Under TE excitation, the reflected signal is ideally TE polarized and routed through the polarizing beam splitter (PBS) before detection.  
    OUT-FPC is adjusted to minimize the signal in the TM port of the PBS, thereby maximizing transmission through the TE port.
    
    A possible concern is the behavior of the guided polarization in the bend.  
    We assume that TE modes follow the bend—remaining in-plane and perpendicular to the local waveguide axis—while TM modes remain out of plane and are unaffected by the bending.  
    We verify this experimentally: after aligning TE input on a straight waveguide as described above, we move the lensed fiber to a bent waveguide and confirm that the reflected signal remains maximized, demonstrating that TE light still reaches the Bragg mirror.  
    The same verification is performed for TM excitation.
    
    \subsection{Fluorescence measurements}
    
    Figures~\ref{figSI:setup}b and \ref{figSI:setup}c show the optical setups used for PL measurements with LWGs and NBs, respectively.  
    As discussed in the main text, the main differences between the two configurations are:
    \begin{itemize}
        \item \textbf{LWG measurements:} A 90:10 beam splitter is used to ensure higher power, a tunable filter (WL Photonics) selects the ZPL, two AOMs (Aerodiode) act as a shutter, and a PBS separates TE and TM polarizations in the detection path.  
        \item \textbf{NB measurements:} A 99:1 beam splitter is used together with a high-speed optical switch (Photonwares) to maximize the efficiency. The PBS is replaced by a 50:50 beam splitter to enable Hanbury Brown and Twiss measurements.
    \end{itemize}
    
    Detected photon counts are processed using a TimeHarp module synchronized with the excitation pulses and the shutter sequence via a pulse generator (PulseBlaster).  
    The full timing sequence is verified using a fast photodetector connected to an oscilloscope and using the TimeHarp module.

\section{Theoretical model}

    \subsection{Optical modes}
    
    The electric-field polarization $\boldsymbol{\epsilon}(\alpha,\mathrm{mode})$ in the laboratory frame depends on the waveguide orientation angle $\alpha$ and on whether the guided mode is TE or TM.
    
    For the TM mode, the electric field always points out of the plane, and therefore
    \[
    \boldsymbol{\epsilon}(\alpha,\mathrm{TM}) = [0,0,1]
    \]
    for any value of $\alpha$.
    
    For a straight waveguide (or nanobeam) with $\alpha=0$, the TE mode lies in the plane with
    \[
    \boldsymbol{\epsilon}(0,\mathrm{TE}) = [0,1,0].
    \]
    For a generic waveguide orientation $\alpha$, the TE polarization is obtained by rotating the straight-waveguide polarization by the rotation matrix
    \[
    \boldsymbol{\epsilon}(\alpha,\mathrm{TE}) = R_{\mathrm{WG}}(\alpha)\,\boldsymbol{\epsilon}(0,\mathrm{TE}),
    \]
    where
    \[
    R_{\mathrm{WG}}(\alpha) =
    \begin{pmatrix}
    \cos\alpha & \sin\alpha & 0 \\
    -\sin\alpha & \cos\alpha & 0 \\
    0 & 0 & 1
    \end{pmatrix}.
    \]

    \subsection{Strain simulations}
    
\begin{figure*}[htb!]
    \centering
    \includegraphics[]{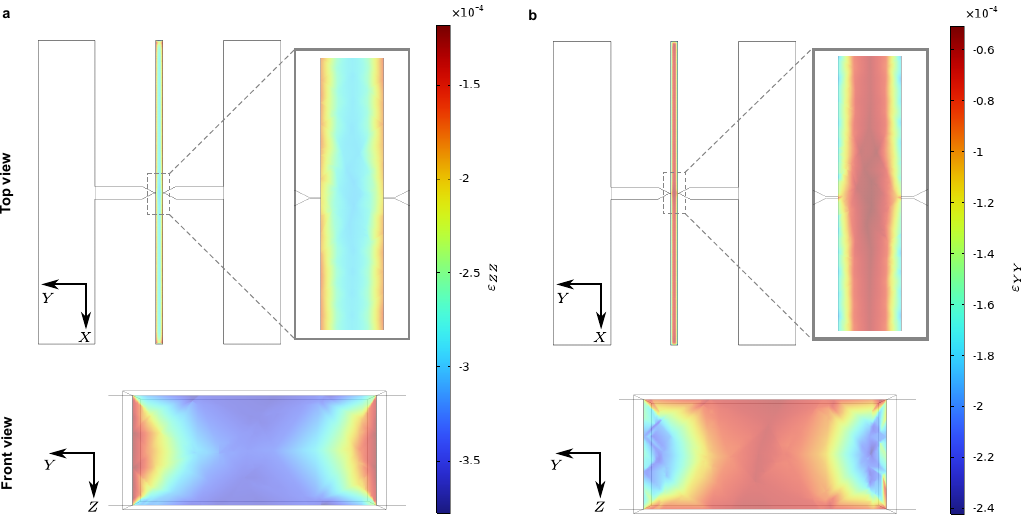}
    \caption{\textbf{Finite-element simulation of strain originating from thermal contraction in one block of the LWG.}
    We simulate a single block including one tether and impose periodic boundary conditions. 
    Panels~\textbf{a} and \textbf{b} show the components $\varepsilon_{ZZ}$ and $\varepsilon_{YY}$, respectively. 
    The coordinate system is defined in the waveguide frame, and the silicon elasticity tensor is transformed accordingly. 
    Results are shown for a waveguide oriented along $\alpha = 0$. 
    Top and front slices through the waveguide center are displayed. 
    The strain distribution is mostly uniform, except near the edges where local variations appear.
    }
    \label{figSI:strain_sims}
    \end{figure*}

    We simulate the strain distribution in a suspended silicon LWG using COMSOL, assuming that the strain originates from thermal contraction upon cooling. 
    Since different waveguides lie along different crystalline directions, the silicon elasticity tensor must be rotated into the corresponding waveguide frame before each simulation (see the coordinate frame in Fig.~\ref{figSI:strain_sims} and Sec.~\ref{sec:elasticity} for details on the tensor transformation).
    
    Figure~\ref{figSI:strain_sims} shows the strain components $\varepsilon_{ZZ}$ (panel~a) and $\varepsilon_{YY}$ (panel~b) obtained from the finite-element simulation of a single LWG block including one tether under periodic boundary conditions.
    
    Because ions may occupy any position within the waveguide cross-section, we compute the spatial averages and standard deviations of each strain component over the entire waveguide domain. 
    These quantities, reported in Table~\ref{tab:strain_sims}, quantify both the dominant strain components and their spatial spread for different LWG orientations.
    
    \begin{table}[htb!]
    \caption{Simulated average strain components $\varepsilon_{ij}$ and their standard deviations over the waveguide for different LWG orientations $\alpha$.}
    \label{tab:strain_sims}
    \begin{tabular}{ccccc}
    \toprule
    $\alpha$ & $0^\circ$ & $43^\circ$ & $-38.5^\circ$ & $90^\circ$ \\
    \midrule
    $\varepsilon_{XX} (\times10^{6})$ & $-7.1 \pm 23.3$ & $-6.9 \pm 23.5$ & $-6.9 \pm 23.5$ & $-7.1 \pm 23.3$ \\
    $\varepsilon_{YY} (\times10^{5})$ & $-9.2 \pm 2.9$ & $-10.7 \pm 3.4$ & $-10.7 \pm 3.3$ & $-9.2 \pm 2.9$ \\
    $\varepsilon_{ZZ} (\times10^{4})$ & $-2.5 \pm 0.4$ & $-2.5 \pm 0.4$ & $-2.5 \pm 0.4$ & $-2.5 \pm 0.4$ \\
    $\varepsilon_{XY} (\times10^{6})$ & $0.003 \pm 6$ & $0.004 \pm 5$ & $0.004 \pm 5$ & $0.003 \pm 6$ \\
    $\varepsilon_{XZ} (\times10^{6})$ & $-0.003 \pm 8$ & $0.23 \pm 5.8$ & $-0.74 \pm 5.9$ & $-0.003 \pm 8$ \\
    $\varepsilon_{YZ} (\times10^{5})$ & $0.001 \pm 3.1$ & $-0.002 \pm 3.1$ & $-0.002 \pm 3.1$ & $0.001 \pm 3.1$ \\
    \botrule
    \end{tabular}
    \end{table}
    First, we note that $\alpha = 0^\circ$ and $\alpha = 90^\circ$ yield identical results, as expected from the symmetry of silicon. 
    The dominant strain component is $\varepsilon_{ZZ}$, whose value is essentially independent of the waveguide orientation $\alpha$. 
    The component $\varepsilon_{YY}$ is also significant, although it shows a somewhat larger spatial variation. 
    Deviations from the mean occur mainly near the waveguide edges, where geometric discontinuities introduce local strain concentrations. 
    The shear components average to nearly zero but exhibit relatively large local fluctuations—which remain negligible compared to the dominant components and are likewise confined to the edges. 
    Such spatial variations may contribute to spectral broadening, but they do not induce an overall frequency shift.

    \subsection{From strain in the waveguide frame to stress in the crystal frame}
    \label{sec:elasticity}
    
    We start from the strain tensor $\varepsilon'$ expressed in the waveguide frame, as this is the output of the COMSOL simulations. 
    To use this strain in our model, we must convert it into the corresponding stress tensor $\sigma$ in the crystal frame.
    
    The elasticity tensor of silicon in the crystal frame (cubic symmetry) is, in Voigt notation,
    \begin{equation}
        C_V = 
        \begin{pmatrix}
            C_{11} & C_{12} & C_{12} & 0      & 0      & 0 \\
            C_{12} & C_{11} & C_{12} & 0      & 0      & 0 \\
            C_{12} & C_{12} & C_{11} & 0      & 0      & 0 \\
            0      & 0      & 0      & C_{44} & 0      & 0 \\
            0      & 0      & 0      & 0      & C_{44} & 0 \\
            0      & 0      & 0      & 0      & 0      & C_{44}
        \end{pmatrix}
        =
        \begin{pmatrix}
            165.7 & 63.9 & 63.9 & 0    & 0    & 0 \\
            63.9  & 165.7 & 63.9 & 0   & 0    & 0 \\
            63.9  & 63.9 & 165.7 & 0   & 0    & 0 \\
            0     & 0    & 0    & 79.6 & 0    & 0 \\
            0     & 0    & 0    & 0    & 79.6 & 0 \\
            0     & 0    & 0    & 0    & 0    & 79.6
        \end{pmatrix}
        \, \text{GPa.}
    \end{equation}
    
    To express the elasticity tensor in the waveguide coordinate frame, we change basis using
    \begin{equation}
        R = R_{\mathrm{WG}}(\alpha)\, R^{\dagger}_{\mathrm{lab}\rightarrow \mathrm{xtal}} .
    \end{equation}
    
    The stiffness matrix is first converted from Voigt notation ($C_V$) to its fourth-order tensor representation ($C_T$) via the Voigt–tensor correspondence.  
    The frame rotation is then applied using Einstein summation,
    \begin{equation}
        C'_{T,ijkl} = R_{im}\, R_{jn}\, R_{ko}\, R_{lp}\, C_{T,mnop}.
    \end{equation}
    Finally, the rotated tensor $C'_T$ is mapped back to Voigt notation as $C'_V$.
    
    The stress in the waveguide frame follows from generalized Hooke’s law,
    \begin{equation}
        \sigma'_{ij} = C'_{T,ijkl}\, \varepsilon'_{k\ell}.
    \end{equation}
    
    To express this stress in the crystal frame, we rotate it back:
    \begin{equation}
        \sigma = R^{\dagger}\, \sigma' \, R .
    \end{equation}

    \subsection{Colormap asymmetry originating from strain}
\begin{figure*}[htb!]
    \centering
    \includegraphics[]{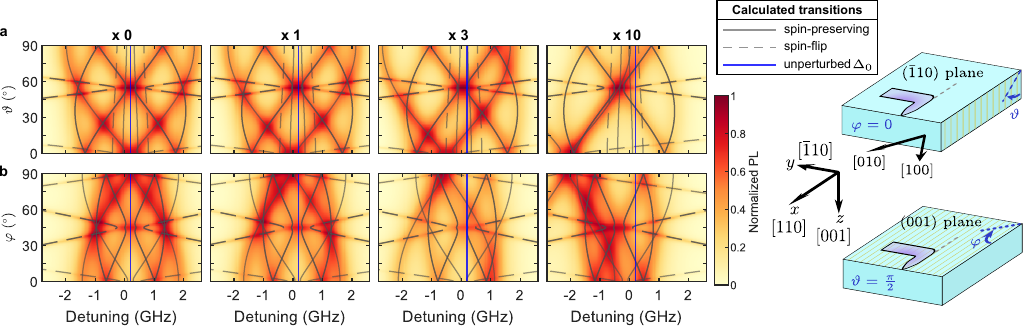}
    \caption{\textbf{Simulations of normalized PL colormaps for different magnitudes of strain.} 
    As in Fig.~2 of the main text, we calculate PL spectra for different magnetic-field orientations. 
    Here we consider a LWG with $\alpha = 90^\circ$ and perform field rotations in the $(\bar{1}10)$ plane (panel~\textbf{a}) and in the $(001)$ plane (panel~\textbf{b}). 
    Using the best-fit parameters, we vary the simulated strain matrix by the multiplicative factors indicated above each colormap. 
    We overlay the predicted spin-preserving (solid lines) and spin-flip (dashed lines) transition frequencies. 
    When strain is absent, the colormap is symmetric around $\Delta_0 = 0.19$~GHz (blue lines), whereas the asymmetry discussed in the main text appears once strain is introduced. 
    This asymmetry becomes increasingly pronounced for larger strain magnitudes.
    }
    \label{figSI:strain_vs_no_strain}
    \end{figure*}

    In the main text, we discuss the asymmetry that appears in the PL spectra when strain is present. 
    Here, we further illustrate this effect by varying the strain amplitude in the model. 
    In Fig.~\ref{figSI:strain_vs_no_strain}, the panel labeled “$\times 0$” shows the expected pattern in the absence of strain: the traces are perfectly symmetric around a central frequency. 
    A slight asymmetry emerges when the strain extracted from our simulations is included (“$\times 1$”). 
    This asymmetry becomes increasingly pronounced for larger strain magnitudes (“$\times 3$” and “$\times 10$”). 
    The effect arises because the different subsites respond differently to strain: depending on their orientation, the transition frequencies shift strongly, weakly, or almost not at all.
    
    \subsection{Coupling strength tunability}
    
    We characterize the tunability of the coupling strength using the magnetic-field orientation under different $g$-tensor conditions, and in doing so we revisit Supplementary Note~2.5 of Ref.~\cite{Raha2020}. 
    That work states that if the ground- and excited-state $g$-tensors are isotropic (or equal), one can always orient the magnetic field such that $g_\perp = 0$, because the relevant $2\times2$ block of the coupling matrix is normal and therefore unitarily diagonalizable.
    
    While mathematically correct, this statement is physically misleading.  
    For isotropic $g$-tensors, the unitary transformation that diagonalizes the relevant $2\times2$ block must also diagonalize the ground- and excited-state spin Hamiltonians.  
    This transformation corresponds to a rotation of the spin quantization axis induced by the applied magnetic field, which spans only a two-dimensional subset of $SU(2)$.  
    In contrast, the unitary needed to diagonalize the coupling block may require an additional degree of freedom: a relative phase rotation about the quantization axis, which cannot be generated by varying the direction of $\mathbf{B}$.
    
    A simple counterexample illustrates this limitation.  
    Consider the coupling operator obtained from the dipole operator projected along the electric-field direction:
    \[
    \hat{M} =
    \begin{pmatrix}
    0_{2\times2} & \hat{\sigma}_x\\[4pt]
    \hat{\sigma}_x & 0_{2\times2}
    \end{pmatrix}, \qquad
    \hat{\sigma}_x = 
    \begin{pmatrix}
    0 & 1\\
    1 & 0
    \end{pmatrix}.
    \]
    Setting $g_\perp = 0$ corresponds to finding a basis in which $\hat{\sigma}_x$ is diagonal.  
    This requires the unitary
    \[
    U = e^{-i\frac{\pi}{4}\sigma_z},
    \]
    which is purely a phase rotation about the $z$ axis.  
    However, such a transformation cannot arise from any spin Hamiltonian of the form $\mathbf{B}\cdot\hat{\mathbf{S}}$: it does not map $\sigma_z$ to a rotated spin operator associated with a real magnetic field direction.  
    Therefore, although $\hat{M}$ is mathematically unitarily diagonalizable, the required unitary is not physically realizable with magnetic-field rotations alone, and $g_\perp$ cannot be forced to zero in this case.\\
    
    Moreover, Ref.~\cite{Raha2020} suggests that alignment of the ground- and excited-state $g$-tensors facilitates cyclicity tunability. 
    Our results indicate the opposite: when both tensors are strongly anisotropic and have parallel axes, the spin quantization axes are effectively pinned, the dipole orientation becomes rigid, and the tunability is minimal, as in the case of Er:Si. 
    In contrast, a finite misalignment allows the two quantization axes to respond differently to $\mathbf{B}$, making it easier to minimize $|g_{\perp}|$ and to reach high cyclicity, as observed for Er:YSO.

    \begin{figure*}[htb!]
    \centering
    \includegraphics[]{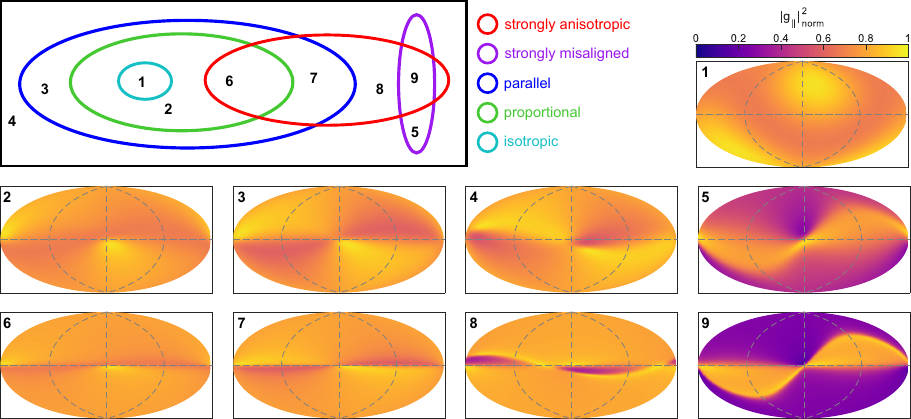}
    \caption{\textbf{Simulations of coupling-strength colormaps for different types of $\boldsymbol{g}$-tensors.} 
    Each set corresponds to a different class of $g$-tensor configurations: red = both tensors anisotropic; purple = strongly misaligned tensor axes; blue = parallel tensor axes; green = proportional tensors ($g_\mathrm{e} \propto g_\mathrm{g}$); cyan = both tensors isotropic. 
    The colormaps are constructed in the same way as in Fig.~3d of the main text and provide representative examples for each case indicated by the numbering in the sets. 
    In all cases we use $(\mathbf{d}_{\parallel}(\mathbf{B}_0),\, \mathbf{d}_{\perp}(\mathbf{B}_0))$ as in our system and assume a cavity mode polarized along $y$. 
    Only the form of the ground- and excited-state $g$-tensors is varied to explore the tunability of the coupling strengths across different types of systems. 
    Case \textbf{1} uses isotropic tensors; cases \textbf{2–5} use mildly anisotropic tensors; cases \textbf{6–9} use strongly anisotropic tensors. 
    In cases \textbf{4} and \textbf{8}, the principal axes of the two tensors are misaligned by approximately $15^\circ$, while in cases \textbf{5} and \textbf{9} they are misaligned by $45^\circ$. 
    Case \textbf{8} is representative of Er:YSO~\cite{Raha2020}, while case \textbf{7} is representative of Er:Si.
    }
    \label{figSI:coupling_tunability}
    \end{figure*}
    
    We further explore this behavior with numerical simulations for several representative forms of ground- and excited-state $g$-tensors. 
    The results, summarized in Fig.~\ref{figSI:coupling_tunability}, show that the tunability is determined by how the two $g$-tensors shape—and constrain—the evolution of the quantization axes:
    \begin{itemize}
        \item \textbf{Isotropic tensors (\textbf{1}) or proportional tensors (\textbf{2}, \textbf{6}).} 
        If both tensors are isotropic (high-symmetry system) or proportional ($g_\mathrm{e} \propto g_\mathrm{g}$), the quantization axes remain aligned for any $\mathbf B$, leading to small tunability. 
        In this situation, the ground- and excited-state spin Hamiltonians are diagonalized by the same rotation, so the dipole operator undergoes only a constrained global rotation. 
        As a result, the relative transition strengths vary little and the coupling tunability remains limited.
    
        \item \textbf{Parallel but non-proportional tensors (\textbf{3}, \textbf{7}).}  
        When the tensors share axes but have different anisotropy ratios (\textbf{3}), the quantization axes are close but not identical, allowing for moderate tunability.  
        If the tensors are strongly anisotropic (\textbf{7}; Er:Si), the quantization axes are pinned to the dominant principal direction except near the plane perpendicular to it, resulting in only narrow regions of tunability.
    
        \item \textbf{Mild anisotropy with slight misalignment (\textbf{4}).}  
        This case behaves similarly to \textbf{3}, with a moderate and relatively broad tunability region.
    
        \item \textbf{Strong anisotropy with slight misalignment (\textbf{8}; Er:YSO).}  
        A small angular offset between the tensors allows the ground- and excited-state quantization axes to respond differently to $\mathbf{B}$, enlarging the regions where $g_\perp$ can be minimized.
    
        \item \textbf{Strongly misaligned tensors (\textbf{5}, \textbf{9}).}  
        The highest tunability arises in low-symmetry systems where the two tensors are highly misaligned (\textbf{5}).  
        When they are also strongly anisotropic (\textbf{9}), the tunability maps exhibit sharper and more distinct features.
    \end{itemize}

    \subsection{Sanity checks}
    
    Given the complexity of the model, it is easy to introduce errors when implementing it in a script. To prevent this, we perform a series of consistency tests (\emph{sanity checks}) that the code must pass to be considered correct.
    
    Since the model involves multiple basis transformations, a primary verification is that all such operations are implemented consistently. One way to test this is to express the model in two equivalent representations—an \emph{active} and a \emph{passive} picture—that should yield identical physical results.  
    
    In the active picture, all quantities are expressed in the laboratory frame, as described in the main text. This requires applying the symmetry operation to all intrinsic quantities of a given subsite, thereby generating the six magnetic classes of the ensemble, while keeping the external quantities fixed.  
    
    In the passive picture, we instead adopt the local frame of each subsite. The intrinsic quantities are thus identical for all subsites, while the external quantities—magnetic field, optical mode, and the Pauli vector defining the spin Hamiltonian—are anti-transformed by the corresponding symmetry operator.  
    
    For each subsite, the coupling-strength $|g_{\parallel}|^2$ must be identical in both pictures. This condition is verified in the script by reproducing main text Fig.~3d from either picture. Other invariants, such as the eigenfrequencies, are also checked for consistency.
    
\begin{figure*}[htb!]
    \centering
    \includegraphics[]{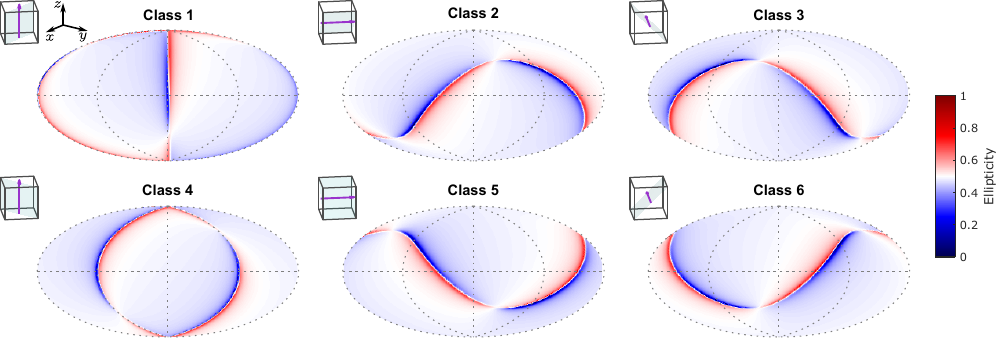}
    \caption{\textbf{Ellipticity of the spin-preserving dipole for different magnetic classes as a function of magnetic-field orientation.} 
    This figure is the analogue of Fig.~3b in the main text, shown here for all six magnetic classes. 
    As expected, the maps are related to one another by the symmetry operations that connect the corresponding classes.
    }
    \label{figSI:ellipticity}
    \end{figure*}

    We further verify that intrinsic quantities of different subsites (e.g., the ellipticity as a function of the magnetic field direction in main text Fig.~3b) are related by the corresponding symmetry operations that map class~1 to the other classes (see Fig~\ref{figSI:ellipticity}).
    
    The correctness of the symmetry operators themselves is validated by visualizing their action on the $C_2$ axis and symmetry planes. Similarly, we check that the operator $R_\mathrm{WG}$ correctly maps the electric field and strain tensors from a straight to a bent waveguide.
    
    A delicate aspect of the model is the repeated diagonalization of the spin Hamiltonian for different magnetic fields and for both the ground and excited states. We ensure that the global phase of the eigenvectors remains consistent across all cases; otherwise, discontinuous phase jumps in the numerical solver would lead to discrepancies between the active and passive pictures.

\section{Ensemble measurements and global fit}

    \subsection{Procedure to compare data and model}\label{SI:fit_compare}
    
    For each pair of angles $(\varphi, \vartheta)$ defining the magnetic-field direction, we compute the eigenfrequencies and the normalized coupling strengths of the four possible transitions to the electric-field mode. 
    For each eigenfrequency, we construct an inhomogeneous broadening profile centered at that frequency with full width at half maximum $\gamma$, where the peak intensity is determined from the PL predicted by the coupling strength (as described in the main text). 
    Summing the contributions from all subsites yields the simulated frequency-sweep spectrum.
    
    After computing the intensity matrix $I$ for each frequency and magnetic-field direction (within a given rotation plane), we normalize it by subtracting the minimum value and dividing by the maximum. 
    An identical normalization procedure is applied to the experimental data, allowing for a direct comparison between model and experiment.
    
    \subsection{Fit parameters}
    We have several parameters in our model.
    We distinguish between parameters that describe the dipoles, parameters for the strain coupling, and general parameters.
    The complex-valued spin-preserving dipole is described by six real parameters:
    \begin{displaymath}
        \mathbf{d}_{\parallel}(\mathbf{B}_0) = \left(
        \begin{array}{c}
             d_{\parallel, x, Re} + i\, d_{\parallel, x, Im}  \\
             d_{\parallel, y, Re} + i\, d_{\parallel, y, Im}  \\
             d_{\parallel, z, Re} + i\, d_{\parallel, z, Im} 
        \end{array}
        \right).
    \end{displaymath}
    Normalization of the intensity to 1 is imposed. The spin-flip dipole is expressed in a similar form:
    \begin{displaymath}
        \mathbf{d}_{\perp}(\mathbf{B}_0) = \sqrt\zeta \cdot \left(
        \begin{array}{c}
             d_{\perp, x, Re} + i\, d_{\perp, x, Im}  \\
             d_{\perp, y, Re} + i\, d_{\perp, y, Im}  \\
             d_{\perp, z, Re} + i\, d_{\perp, z, Im} 
        \end{array}
        \right).
    \end{displaymath}
    We relate the absolute value of $\mathbf{d}_{\perp}(\mathbf{B}_0)$ to the normalized $\mathbf{d}_{\parallel}(\mathbf{B}_0)$ using a flip factor $\zeta$:
    \begin{equation}
        \zeta = \left|
        \frac{\mathbf{d}_{\perp}(\mathbf{B}_0)}
        {\mathbf{d}_{\parallel}(\mathbf{B}_0)}
        \right|^2.
    \end{equation}
    
   Strain is described by three free parameters: $A_1$, $A_2$, $A_3$, as explained in the main text.
   
    Other general parameters that relate to both the dipoles and the strain are 
    \begin{itemize}
        \item     $\gamma$, which describes the linewidth of the inhomogeneous broadening, as defined in Section \ref{SI:fit_compare}.
        \item $\Delta_0$, the center of the inhomogeneous broadening with respect to 194.953~THz (1537.761~nm) in the absence of strain.
        \item $C_B$, a factor to adjust the magnetic field magnitude with respect to our calibration -- we multiply the expected magnetic field by this value.      
    \end{itemize}

    The best-fit parameters are reported in Table~\ref{tabSI:fit}.
    \begin{table*}[htb!]
        \centering
        \small
        \caption{\textbf{Best-fit parameters describing the transition dipoles, strain coupling, and general properties.}
         Values are reported with one–sigma uncertainties. The dipole components are expressed in the laboratory frame, assuming $\mathbf{B}_0$ aligned along $y$.}
        \label{tabSI:fit}
        \begin{tabular}{@{}lcc@{}}
        \toprule
        \textbf{Group} & \textbf{Symbol} & \textbf{Value} \\
        \midrule
        \multirow{6}{*}{$\mathbf{d}_{\parallel}(\mathbf{B}_0)$} 
         & $d_{\parallel,x,Re}$ & $0.165 \pm 0.005$ \\
         & $d_{\parallel,x,Im}$ & $0.040 \pm 0.011$ \\
         & $d_{\parallel,y,Re}$ & $-0.771 \pm 0.005$\\
         & $d_{\parallel,y,Im}$ & $0.136 \pm 0.016$\\
         & $d_{\parallel,z,Re}$ & $-0.292 \pm 0.008$  \\
         & $d_{\parallel,z,Im}$ & $-0.384 \pm 0.008$ \\
        \midrule
        \multirow{7}{*}{$\mathbf{d}_{\perp}(\mathbf{B}_0)$} 
         & $d_{\perp,x,Re}$ & $0.023 \pm 0.014$ \\
         & $d_{\perp,x,Im}$ & $-0.577 \pm 0.006$ \\
         & $d_{\perp,y,Re}$ & $0.344 \pm 0.013$ \\
         & $d_{\perp,y,Im}$ & $-0.396 \pm 0.013$  \\
         & $d_{\perp,z,Re}$ & $-0.244 \pm 0.011$ \\
         & $d_{\perp,z,Im}$ & $0.071 \pm 0.013$ \\
         & $\zeta$ & $0.607 \pm 0.006$ \\
        \midrule
        \multirow{3}{*}{\textbf{Strain}} 
         & $A_{1}$ & $3.34 \pm 0.48$~Hz/Pa \\
         & $A_{2}$ & $-0.50 \pm 0.50$~Hz/Pa \\
         & $A_{3}$ & $-7.30 \pm 0.21$~Hz/Pa\\
        \midrule
        \multirow{3}{*}{\textbf{General}} 
         & $\gamma$ & $0.686 \pm 0.003~\text{GHz}$ \\
         & $\Delta_{0}$ & $0.192 \pm 0.051~\text{GHz}$ \\
         & $C_{B}$ & $0.8416 \pm 0.0005$ \\
        \bottomrule
        \end{tabular}
    \end{table*}

\begin{figure*}[htb!]
    \centering
    \includegraphics[]{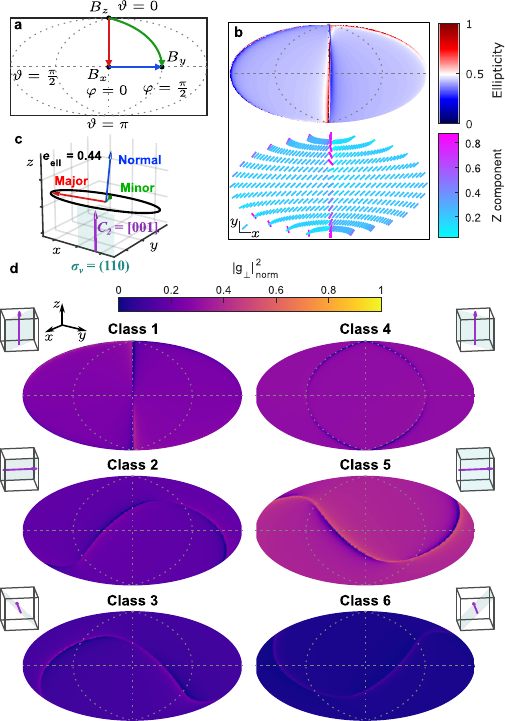}
    \caption{\textbf{Calculated dipole properties as a function of magnetic field orientation for the spin-flip transition.}
    \textbf{a.} Hammer projection of the magnetic-field orientation in the laboratory frame.
    \textbf{b.} Dipole polarization properties for a class~1 ion. Top: ellipticity $e_{\mathrm{ell}}$, ranging from 0 (linear) to 1 (circular). Bottom: vector-field map of the major-axis direction projected onto the $xy$-plane; the colormap encodes its $z$ component, which is positive by normalization.
    \textbf{c.} Mean polarization ellipse ($e_{\mathrm{ell}} = 0.44$) obtained by averaging the major, minor, and normal axes over all magnetic-field orientations, excluding a $10^\circ$ region around the $zx$-plane to capture typical behavior. Red, green, and blue arrows indicate the mean major, minor, and normal axes, with directions $[-0.71, -0.67, 0.23]$, $[0.63, -0.50, 0.59]$, and $[-0.29, 0.58, 0.76]$ in the laboratory frame.
    \textbf{d.} Normalized squared coupling strength $|g_{\perp}|_{\mathrm{norm}}^{2}$, obtained by dividing $g_{\perp}$ by $|\mathbf{E}|\,|\mathbf{d}_{\parallel}|$, as for $g_{\parallel}$, for the spin-flip dipole coupled to a NB optical mode polarized along $y$.
    Each panel corresponds to the specified magnetic class, oriented as shown in the schematics alongside the map.
    The overall angular dependence follows the same trends as in Fig.~3 of the main text, but with different magnitudes.
    }
    \label{ext_fig_flip}
    \end{figure*}

    For completeness, in Fig.~\ref{ext_fig_flip} we report the calculated dipole properties resulting from the best-fit parameters for the spin-flip transition.
    %FigE4: same as Fig3 for spin-flip transition

    \subsection{Fit routine}
    
    The model contains several parameters, grouped into three categories. 
    The dipole parameters determine the relative intensity of each transition, whereas the strain and the general parameters determine the eigenfrequencies and the shape of the inhomogeneous broadening. 
    Because dipole and strain parameters influence the model in fundamentally different ways, we adopt an iterative fitting procedure. 
    We first fix the strain parameters to reasonable initial values and optimize all remaining parameters. 
    We then fix the optimized dipole parameters and refine the strain-related parameters. 
    This alternation is repeated for several cycles until all parameters converge.
    
    The optimization is performed using the MATLAB function \texttt{fmincon}, minimizing the reduced $\chi^2$ between model and data (as described in the main text). 
    
    To reduce the risk of converging to local minima, we employ a multi-stage strategy:
    \begin{enumerate}
        \item a \emph{GlobalSearch} run using \texttt{fmincon} with gradient information,
        \item a \emph{MultiStart} run using \texttt{fmincon} from many randomized initial conditions,
        \item a final local refinement using \texttt{fmincon}.
    \end{enumerate}
    GlobalSearch and MultiStart explore the parameter space using many different starting points, increasing the likelihood of locating the global minimum, while the final refinement ensures precise convergence of the optimal solution.

    \subsection{Magnetic field sources of uncertainty}
    In the magnetic-field rotation measurements, the dominant source of uncertainty arises from the orientation of the applied magnetic field. 
    We estimate the field direction based on the calibration of our homemade vector magnet~\cite{DaPrato2025}, although this calibration is reliable only to within approximately $10\%$. 
    In addition, stray magnetic fields in the environment are not accounted for. 
    For example, a small permanent magnet located roughly 5~cm above the sample generates a background field of order 1~mT at the sample position. 
    Devices not located precisely at the center of the vector magnet also experience a slightly distorted magnetic field.
    
    The current source used to drive the coils (Rigol DP832A) introduces an additional source of error. 
    It has a resolution of 1~mA and an accuracy of $0.2\% + 5$~mA. 
    Given calibration values up to 150~mT/A, this uncertainty becomes non-negligible for the single-ion measurements, where the magnetic fields used are much smaller than those applied in the ensemble measurements. 
    This helps explain why sharp features in the lifetime dependence on magnetic-field orientation are difficult to resolve.
    
    Because of the combination of these uncertainties, it is challenging to assign a rigorous error bar to the ensemble measurements. 
    We therefore adopt a phenomenological angular uncertainty of $3^\circ$, which corresponds approximately to the angular resolution of the colormaps. 
    This value is not used in the fitting routine itself but only in post-processing to estimate the final metrics and the uncertainties on the best-fit parameters.

    \subsection{Additional transition in the background}
    \begin{figure*}[htb!]
    \centering
    \includegraphics[]{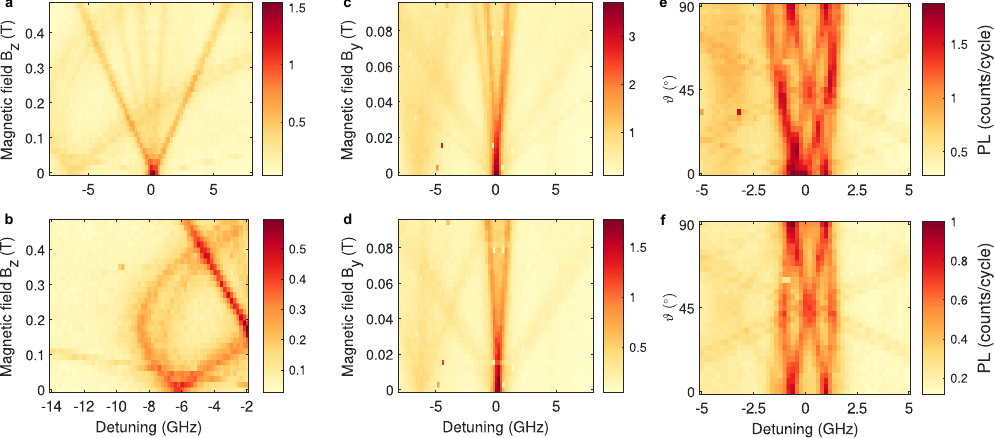}
    \caption{\textbf{Additional transition observed in the background of the ensemble measurements.} 
    The excitation frequency is swept around 1537.761~nm (site A) while a 0.5~nm filter centered at the same wavelength selects the detected PL.
    \textbf{a.} Magnetic-field sweep along the $z$ direction. TM mode in input and output.  
    \textbf{b.} Same measurement as in \textbf{a}, zoomed in on the faint transition around $-6.2$~GHz at zero magnetic field.  
    \textbf{c.} Magnetic-field sweep along the $y$ direction. TE mode in input and output.  
    \textbf{d.} Same measurement as in \textbf{c} but with TM polarization in output.  
    \textbf{e.} Magnetic-field rotation in the $(001)$ plane, with the field magnitude fixed at 94~mT. TE mode in input and output.  
    \textbf{f.} Same measurement as in \textbf{e} but with TM polarization in output.}
    \label{figSI:background_transition}
    \end{figure*}
    
    The background observed in the ensemble measurements is not flat because of fluorescence from an additional optical transition (Fig.~\ref{figSI:background_transition}e,f).  
    Figures~\ref{figSI:background_transition}a–d show that, in the absence of a magnetic field, this transition is centered around $-6.2$~GHz relative to the center of the site~A inhomogeneous broadening.  
    When a magnetic field is applied along the $z$ direction (up to 0.5~T), the Zeeman splitting of this faint transition becomes strongly nonlinear.  
    The transition is also polarization dependent, as seen by comparing Fig.~\ref{figSI:background_transition}c with Fig.~\ref{figSI:background_transition}d, and Fig.~\ref{figSI:background_transition}e with Fig.~\ref{figSI:background_transition}f.
    
    We speculate that this additional fluorescence originates from ions belonging to site~P~\cite{Gritsch2022}.  
    Site~P exhibits fully lifted crystal-field degeneracy even at zero magnetic field, producing several closely spaced manifold levels, which would explain the nonlinear Zeeman response.  
    Moreover, site~P is believed to originate from precipitate-related defects and is therefore not fully reproducible; some of its transitions lie near the site~A ZPL, but their exact frequencies are not known a priori. Analysis of the time-resolved PL yields a lifetime of $1.13(5)$~ms for this transition—significantly longer than the lifetime of site~A, as expected.
   
    \subsection{Power sweeps with an ensemble}
    \label{sec:power_sweeps_ensemble}
    % best fit parameters: 
    % TETE I0 = 1.5, P_sat = 30 nW, A1 = 3.3e5(nW)^-1, N = 5.0e-3
    % TETM I0 = 0.75, P_sat = 32 nW, A1 = 1.7e5(nW)^-1 , N = 5.0e-3
    % TMTE I0 = 1.5, P_sat = 75 nW, A1 = 1.7e5(nW)^-1, N = 4.5e-3
    % TMTM I0 = 0.75, P_sat = 66 nW, A1 = 8.4e4(nW)^-1, N = 4.7e-3
    
    \begin{figure*}[htb!]
    \centering
    \includegraphics[]{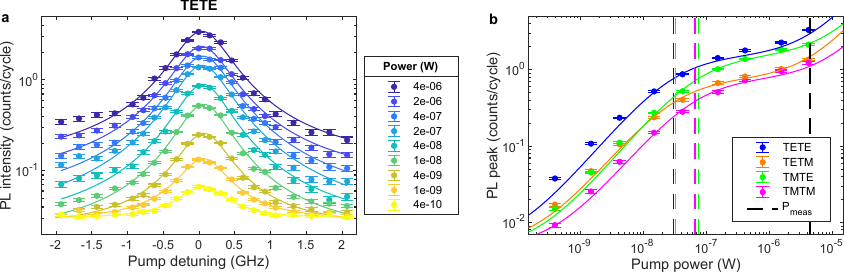}
    \caption{\textbf{Power sweeps for an ensemble of ions.}
    In the labels TETE, TETM, TMTE, and TMTM, the first two letters refer to the input polarization mode (TE or TM) and the last two to the detected output polarization. 
    \textbf{a.} PL spectra recorded at different excitation powers, together with Lorentzian fits for the TETE configuration. 
    \textbf{b.} Extracted PL peak intensity (after background subtraction) as a function of the excitation power for all four polarization combinations. 
    Solid lines show the best fits to Eq.~(\ref{eq:power_ensemble}), and dashed vertical lines mark the corresponding saturation powers. 
    The black dashed line indicates the excitation power used for the ensemble measurements discussed in the main text.}
    \label{figSI:power_sweep_ensemble}
    \end{figure*}
    
    We measure the PL intensity of an ensemble of ions in a LWG as a function of the optical pump power. 
    All measurements are performed in the absence of an external magnetic field, with the possibility to select either TE or TM polarization at both the input and output, as described in Section \ref{sec:setup}. 
    The notation TETE, for example, indicates excitation in TE mode and detection in TE mode; similarly, TETM, TMTE, and TMTM correspond to all other combinations of input and output polarizations.
    
    For each power value, the emission spectrum is fitted with a Lorentzian function:
    \begin{equation}\label{eq:lorentzian}
        \mathrm{PL}(\Delta) = \frac{A}{(\Delta - \Delta_0)^2 + B} + C,
    \end{equation}
    where $\Delta_0$ is the central frequency, $A$ the amplitude, $B$ the width parameter, and $C$ a constant background. 
    From the fit we extract the peak intensity after background subtraction, $I = A/B$. 
    
    Figure~\ref{figSI:power_sweep_ensemble}a shows the fitted spectra for the TETE configuration. 
    Figure~\ref{figSI:power_sweep_ensemble}b reports the extracted peak intensity $I$ as a function of the instantaneous excitation power $P$ for the four input–output polarization combinations. 
    The data are fitted using the model \cite{Buzzi2025}
    \begin{equation}\label{eq:power_ensemble}
        I(P) = I_0 \left( \frac{P}{P + P_{\mathrm{sat}}} \right) + \frac{I_{0,\,\mathrm{O}}}{P_{\mathrm{sat},\,\mathrm{O}}} P + N,
    \end{equation}
    where $P_{\mathrm{sat}}$ is the saturation power for the considered transition, $N$ accounts for residual background, and the linear term in $P$ represents additional emission channels below saturation. 
    This second term is required because, as seen in Fig.~\ref{figSI:power_sweep_ensemble}b, the PL signal increases again at high power even after the main transition has saturated. 
    
    A plausible explanation is the contribution from the optically active site~O~\cite{Gritsch2022}, whose excitation is independent of the laser wavelength.
    In principle, this contribution should be compensated by the background subtraction ($C$ in the fit); however, Fig.~\ref{figSI:power_sweep_ensemble}a shows that at high power the Lorentzian tails do not align perfectly with the data, and the fitted background level clearly rises with increasing power. 
    This effect cannot be attributed to detector dark counts or stray light.
    
    In any case, the key parameter we extract from the fit is the saturation power $P_{\mathrm{sat}}$ of the site~A optical transition. 
    We obtain $P_{\mathrm{sat}} =$ 30~nW, 32~nW, 75~nW, and 66~nW for the TETE, TETM, TMTE, and TMTM configurations, respectively. 
    As expected (see Section~\ref{sec:saturation}), for excitation powers well above the saturation threshold the input polarization becomes irrelevant.
    
    The ensemble measurements presented in the main text are performed at an excitation power of \SI{4.4}{\micro\watt}, far above the saturation level. 
    From this it follows that the excitation probability is essentially independent of the input polarization, as we will show in the next Section.

    \subsection{Effective excitation power and polarization dependence}
    \label{sec:saturation}
    
    For excitation powers well above the saturation threshold of the site~A transition ($P / P_{\mathrm{sat}} \gg 1$) but still below that of other optically active defects ($P / P_{\mathrm{sat},\,\mathrm{O}} \ll 1$), Eq.~(\ref{eq:power_ensemble}) simplifies to
    \begin{equation}
        I \approx I_0 + \frac{I_{0,\,\mathrm{O}}}{P_{\mathrm{sat},\,\mathrm{O}}} P + N.
    \end{equation}
    
    The effective optical power available to excite the ions depends on the overlap between the electric-field polarization of the guided mode and the transition dipole moment. 
    For excitation through the TE mode, the effective power is
    \begin{equation}
        P_{\mathrm{TE}} = P \frac{|{\boldsymbol{\epsilon}}_{\mathrm{TE}} \cdot \mathbf{d}|^2}{|\mathbf{d}|^2} \leq P,
    \end{equation}
    where ${\boldsymbol{\epsilon}}_{\mathrm{TE}}$ is the electric-field unit vector of the TE mode and $\mathbf{d}$ is the transition dipole vector. 
    
    For excitation through the TM mode (${\boldsymbol{\epsilon}}_{\mathrm{TM}}$), the effective power is further reduced by a factor of $1/2$ because of the fully transparent mirror of the LWG:
    \begin{equation}
        P_{\mathrm{TM}} = P \cdot \frac{1}{2} \cdot \frac{|{\boldsymbol{\epsilon}}_{\mathrm{TM}} \cdot \mathbf{d}|^2}{|\mathbf{d}|^2} \leq \frac{1}{2}P.
    \end{equation}
    The additional emission channel associated with site~O is expected to be largely polarization-independent, as it does not rely on resonant excitation of a specific transition but instead scales with the total optical power.
    
    These considerations explain the different saturation powers observed for TE and TM excitation in Fig.~\ref{figSI:power_sweep_ensemble}b, with the TM input requiring approximately twice the power of the TE input to reach saturation.
    In the present measurements, no external magnetic field is applied, so the dipole orientation is unknown and appears to project similarly onto both TE and TM modes, as inferred from Fig.~\ref{figSI:power_sweep_ensemble}b. 
    In general, when a magnetic field is applied, the dipole assumes a different orientation; however, since the excitation powers used in our experiments are pretty high, the actual dipole orientation is mostly irrelevant, as even a small projection of the dipole onto the considered mode field is sufficient to reach the saturation regime.

    \subsection{Comparison between TE and TM excitation}
    
\begin{figure*}[htb!]
    \centering
    \includegraphics[]{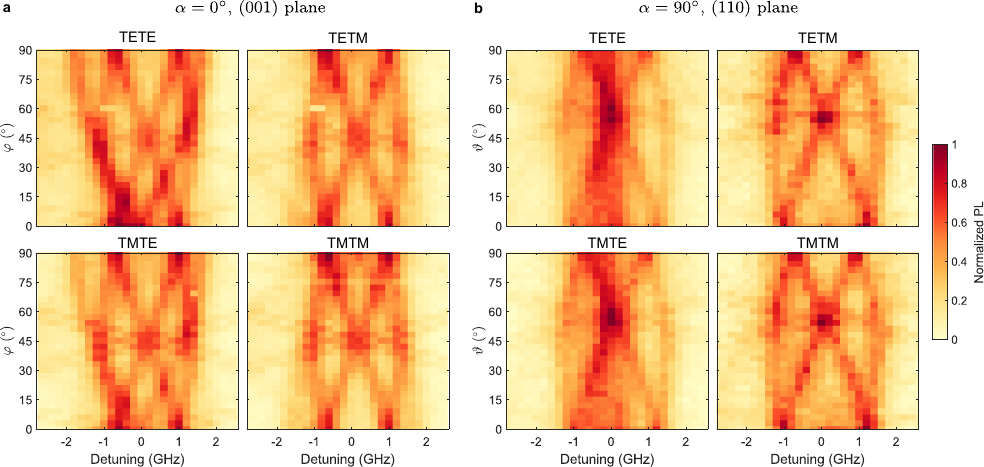}
    \caption{\textbf{Magnetic-field-dependent PL spectra measured on an ensemble for different combinations of input and output polarization modes.}
    The PL spectra are normalized by subtracting the background (minimum signal) and dividing by the maximum intensity. 
    The horizontal axis represents the laser detuning from 1537.761~nm, corresponding to the site~A center reported in Ref.~\cite{Gritsch2022}. 
    The labels TETE, TETM, TMTE, and TMTM denote the combinations of input and output polarization modes, where the first two letters refer to the excitation (TE or TM) and the last two to the detected polarization. 
    \textbf{a.} Magnetic-field rotation in the $(001)$ plane, with $\vartheta = \pi/2$ and $\varphi$ varied from 0 to $\pi/2$, at a fixed field magnitude of 94~mT. The LWG orientation is $\alpha = 0^\circ$. 
    \textbf{b.} Magnetic-field rotation in the $(110)$ plane, with $\varphi = \pi/2$ and $\vartheta$ varied from 0 to $\pi/2$, at a fixed field magnitude of 94~mT. The LWG orientation is $\alpha = 90^\circ$. 
    The similarity between the first and second rows is consistent with the model discussed in the high-power excitation regime.}
    \label{figSI:TE_vs_TM}
    \end{figure*}

    To experimentally verify the predictions of the saturation model, we measure the magnetic-field-dependent PL spectra, as discussed in Fig.~2 of the main text, this time using the TM mode for excitation. 
    Figure~\ref{figSI:TE_vs_TM} shows the four possible combinations of input and output polarization modes for two representative cases: 
    $\alpha = 0^\circ$ (panel~a), with magnetic-field rotation in the $(001)$ plane, and 
    $\alpha = 90^\circ$ (panel~b), with rotation in the $(110)$ plane. 
    
    The first row corresponds to TE excitation, while the second row corresponds to TM excitation. 
    As expected, the two rows appear very similar, confirming that the input polarization has little influence in the high-power excitation regime. 
    In contrast, the detected (output) polarization plays a crucial role: the spectra show clear differences between TE and TM detection, consistent with the polarization filtering of the collected emission.
        
    \subsection{Datasets included in the analysis}

\begin{figure*}[htb!]
    \centering
    \includegraphics[]{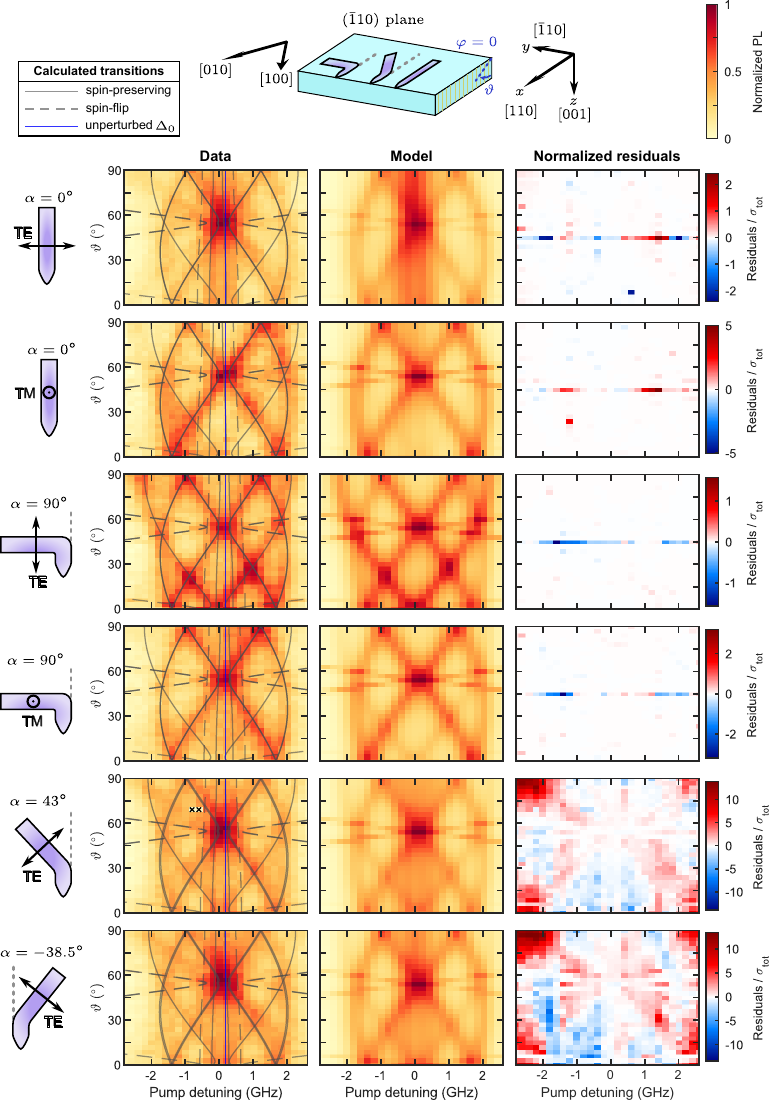}
    \caption{\textbf{Ensemble PL excitation spectra for magnetic-field rotation in the $\boldsymbol{(\bar{1}10)}$ plane.}
    The field orientation is defined by $\varphi = 0$, with $\vartheta$ swept from $0$ to $\pi/2$ at a fixed magnitude of 125~mT. All colormaps and normalization conventions are defined as in Fig.~2c of the main text, with the addition of normalized residual maps to compare experiment and model. All datasets from the $(\bar{1}10)$ sweep plane included in the global fit are shown.
    }
    \label{figSI:dataset_-110}
    \end{figure*}

\begin{figure*}[htb!]
    \centering
    \includegraphics[]{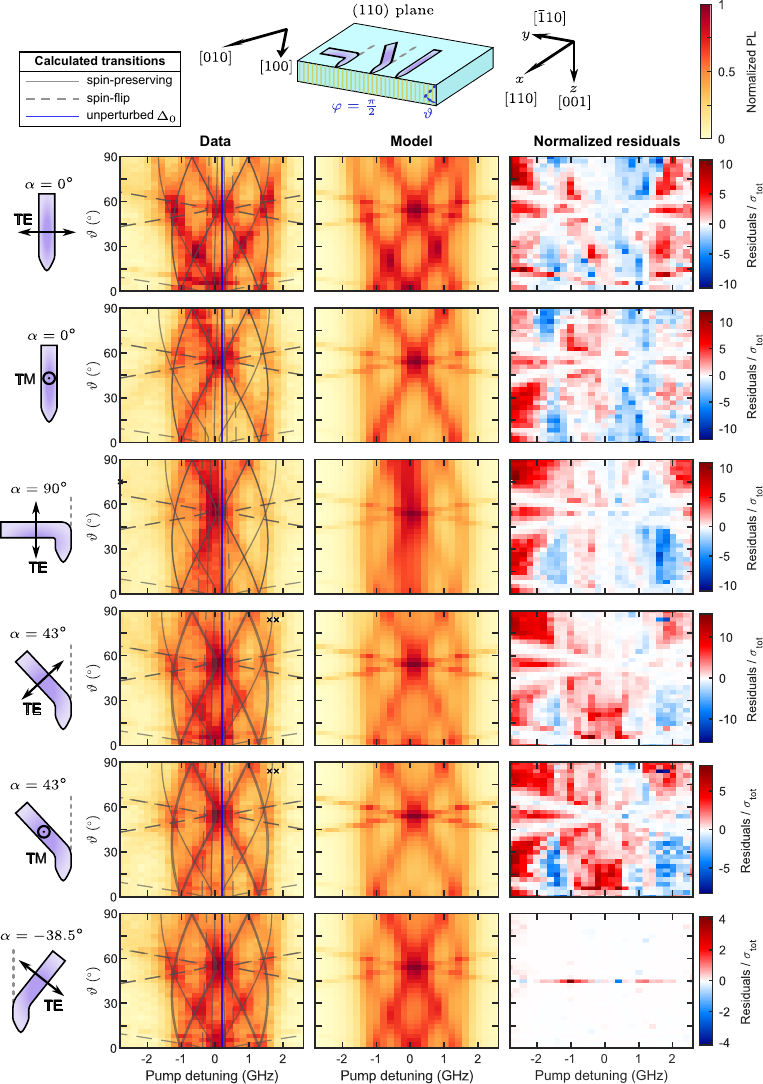}    
    \caption{\textbf{Ensemble PL excitation spectra for magnetic-field rotation in the $\boldsymbol{(110)}$ plane.}
    The field orientation is defined by $\varphi = \pi/2$, with $\vartheta$ swept from $0$ to $\pi/2$ at a fixed magnitude of 94~mT. All colormaps and normalization conventions are defined as in Fig.~2c of the main text, with the addition of normalized residual maps to compare experiment and model. All datasets from the $(110)$ sweep plane used in the global fit are shown.
    }
    \label{figSI:dataset_110}
    \end{figure*}

\begin{figure*}[htb!]
    \centering
    \includegraphics[]{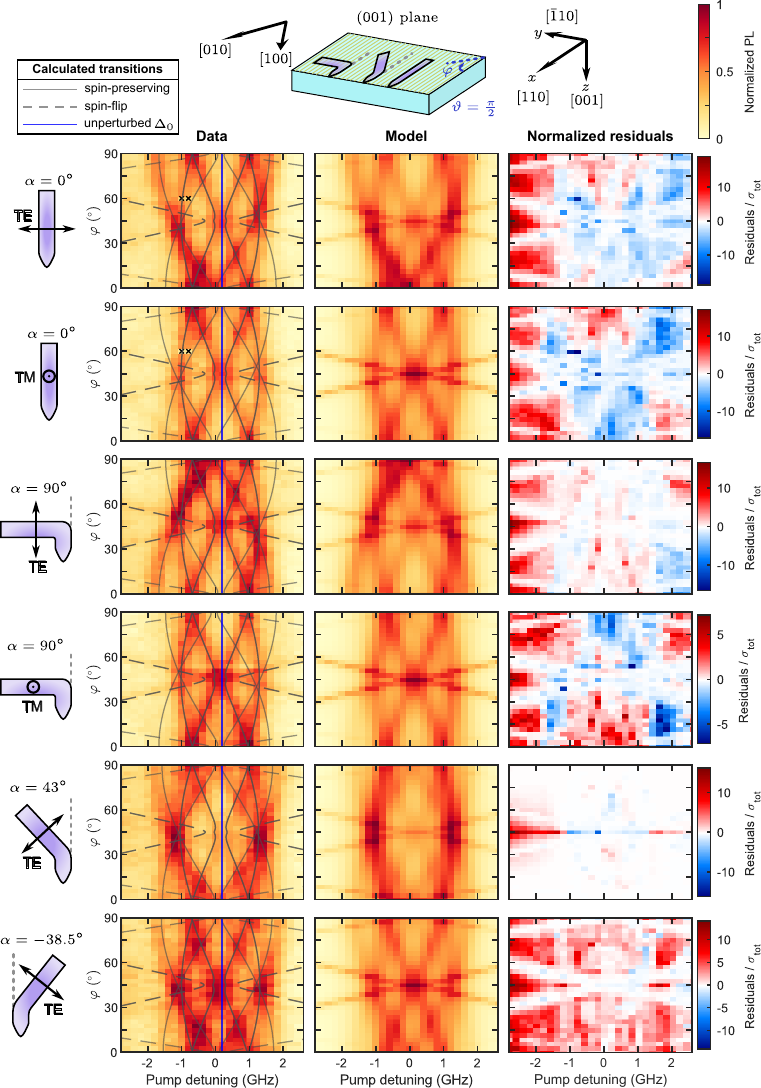}
    \caption{\textbf{Ensemble PL excitation spectra for magnetic-field rotation in the $\boldsymbol{(001)}$ plane.}
    The field orientation is defined by $\vartheta = \pi/2$, with $\varphi$ swept from $0$ to $\pi/2$ at a fixed magnitude of 94~mT. All colormaps and normalization conventions are defined as in Fig.~2c of the main text, with the addition of normalized residual maps to compare experiment and model. All datasets from the $(001)$ sweep plane used in the global fit are shown.
    }
    \label{figSI:dataset_001}
    \end{figure*}

\begin{figure*}[htb!]
    \centering
    \includegraphics[]{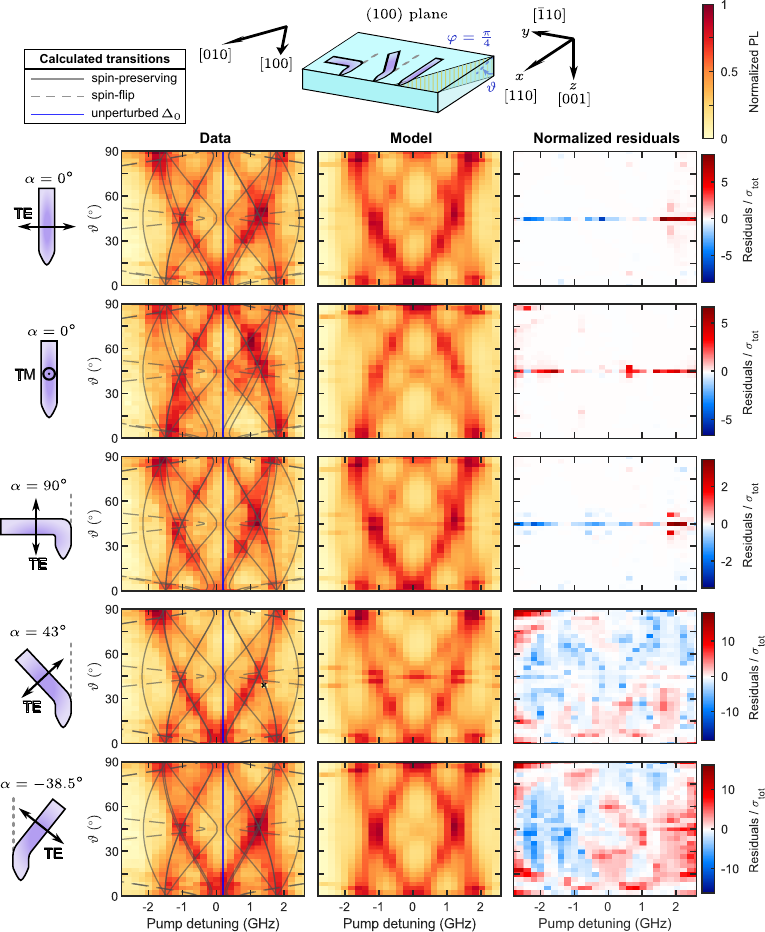}
    \caption{\textbf{Ensemble PL excitation spectra for magnetic-field rotation in the $\boldsymbol{(100)}$ plane.}
    The field orientation is defined by $\varphi = \pi/4$, with $\vartheta$ swept from $0$ to $\pi/2$ at a fixed magnitude of 133~mT. All colormaps and normalization conventions are defined as in Fig.~2c of the main text, with the addition of normalized residual maps to compare experiment and model. All datasets from the $(100)$ sweep plane used in the global fit are shown.
    }
    \label{figSI:dataset_100}
    \end{figure*}

    The full dataset consists of measurements performed on four LWGs with orientation angles
    $\alpha = 0^\circ$, $90^\circ$, $43^\circ$, and $-38.5^\circ$.
    For each waveguide, the magnetic field is swept in four crystallographic planes:
    $(\bar{1}10)$ (Fig.~\ref{figSI:dataset_-110}), $(110)$ (Fig.~\ref{figSI:dataset_110}),
    $(001)$ (Fig.~\ref{figSI:dataset_001}), and $(100)$ (Fig.~\ref{figSI:dataset_100}).
    The input polarization is TE, while both TE and TM components are recorded at the output. In the analysis, only a subset of the TM-output datasets is included.
    This choice is motivated by two considerations.
    First, the signal-to-noise ratio is frequently lower for TM output, since only approximately half of the guided power is collected in this polarization (as discussed earlier).
    Second, the TM polarization orientation is identical for all waveguide angles; including all TM datasets would therefore disproportionately weight this polarization class and bias the statistical balance of the combined dataset.
    The statistics of the individual datasets included in the fitting procedure are summarized in main text Fig.~8.
    
\section{Measurements on cavity-enhanced ions}

    \subsection{Gas tuning of the cavity resonance} \label{sec:gas_tuning}
\begin{figure*}[htb!]
    \centering
    \includegraphics[]{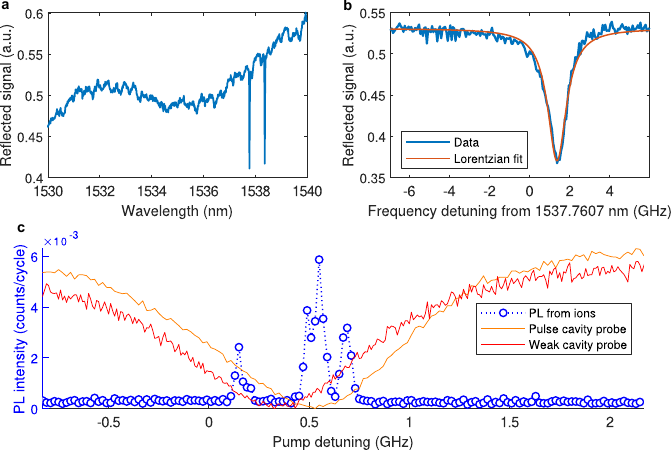}
    \caption{\textbf{Cavity probing with different methods.}
    \textbf{a}. Coarse scan of the two nanobeam cavities (see Fig.~\ref{figSI:device_design}e) of the device under study, acquired after the final PL measurement.
    \textbf{b}. Fine scan of the low-wavelength cavity in \textbf{a}. The profile is fitted with a Lorentzian model (red trace). A few GHz blueshift is observed, likely caused by partial melting of the nitrogen layer during high-power probing. 
    \textbf{c}. Broad PL spectrum centered at 1537.7607~nm. The orange trace shows the normalized reflected pulse, representing the cavity profile in the free-carrier dispersion regime. The red trace shows the normalized intrinsic cavity profile measured using weak-light probing.
    }
    \label{figSI:cavity_res}
    \end{figure*}

    We characterize the devices at room temperature using reflection measurements with the setup described in Sec.~\ref{sec:setup}. 
    As introduced earlier (see Fig.~\ref{figSI:device_design}e), each device consists of two nanobeams designed to be nominally identical, but exhibiting slightly different resonance frequencies of their fundamental TE modes due to fabrication-induced variations. As a result, two distinct cavity resonances are observed for each device.
    We first perform a coarse scan by sweeping the laser piezo to identify these two resonances (e.g., Fig.~\ref{figSI:cavity_res}a).
    We then focus on the cavity of interest and perform a fine scan by locking the laser to the wavelength meter (e.g., Fig.~\ref{figSI:cavity_res}b). 
    The resonance wavelength of the cavity studied here is 1549.50~nm at room temperature, with a linewidth of 0.9~GHz. 
    After cooling below 4~K, thermal effects and the vacuum environment induce a blue shift of about 12.7~nm, placing the resonance at 1536.80~nm.
    
    To match the site~A optical transition (1537.761~nm), the cavity must be red-detuned. 
    We achieve this using nitrogen gas tuning, following an approach similar to Ref.~\cite{Raha2021}. 
    Nitrogen is frozen inside a copper tube thermally anchored to the 4~K plate of the dilution refrigerator. 
    By heating the tube, nitrogen sublimates and subsequently redeposits onto the device, modifying the local refractive index. 
    The magnitude of the cavity shift depends on the thickness of the deposited nitrogen layer. 
    We monitor the cavity resonance in real time during this process to determine when to stop heating. 
    After gas tuning, the relevant nanobeam cavity is centered at 1537.96~nm.
    
    To fine tune the resonance, we use a resonant laser with a power above a threshold ($\gtrsim \SI{1.5}{\micro\watt}$), which removes nitrogen in a controlled manner. 
    This allows us to position the cavity resonance with a resolution of about 200~MHz. 
    After each nitrogen-removal “shot”, we measure the updated cavity position to determine the next step. 
    Once the resonance approaches within approximately 5–10~GHz of the target frequency, we stop using this standard high-power probing technique. 
    Indeed, the minimum optical power required to obtain a measurable reflected signal on a photodiode ($>\SI{0.1}{\micro\watt}$) causes a noticeable thermo-optic redshift~\cite{Barclay2005, Chan2011} and may partially melt the nitrogen layer.
    
    To avoid these effects, we switch to a low-power probing method. 
    We continue performing reflection measurements, but now use extremely weak light ($\sim$~pW) and detect the reflected signal with the SNSPDs (e.g., red trace in Fig.~\ref{figSI:cavity_res}c). 
    In the dataset shown, the final high-power scan is performed at 1537.9153~nm, after which low-power probing is used until we reach 1537.7580~nm, with a linewidth of 1.18~GHz (\ref{figSI:cavity_res}c).
    
    After completing all PL measurements, we perform a standard high-power reflection measurement to assess any additional drift. 
    We find the cavity at 1537.7498~nm, with a linewidth of 1.14~GHz (\ref{figSI:cavity_res}a,b). 
    This small blueshift is likely caused by nitrogen melting during the final probing step.
    
    \subsection{Free-carrier dispersion versus thermo-optic effect}
    
    At very low optical powers, such as those used in the weak-probe measurements, the cavity resonance remains at its nominal frequency. 
    When the power is increased to the nW level—comparable to the excitation pulses used in the fluorescence measurements—the cavity instead experiences a blue shift due to free-carrier dispersion, where photogenerated carriers modify the refractive index~\cite{Barclay2005, Chan2011}. 
    At higher powers, the thermo-optic effect becomes dominant and produces a red shift of the cavity resonance, as observed in the high-power reflection measurements.
    
    In our PL measurements, the cavity operates in the free-carrier-dispersion regime; therefore, its resonance does not coincide with the value measured in the weak-probe scans. 
    However, we can track the actual resonance frequency during the fluorescence cycles. 
    In the time-resolved fluorescence, we integrate over a window that includes the reflected excitation pulse (0–\SI{10}{\micro\second}). 
    Although a shutter blocks most of this pulse, a small leakage remains detectable. 
    Figure~\ref{figSI:cavity_res}c shows the reflected pulse as a function of laser detuning (orange trace), which is shifted by roughly 200~MHz relative to the weak-probe profile (red trace).
    
    Finally, alongside the cavity profiles in the two regimes, we also report a broad spectral scan of the ions, obtained from the same data used to extract the real-time cavity profile. 
    No additional strongly coupled ions are observed beyond those discussed in the main text.
    
    \subsection{Cavity and ion wandering}

    \begin{figure*}[htb!]
    \centering
    \includegraphics[]{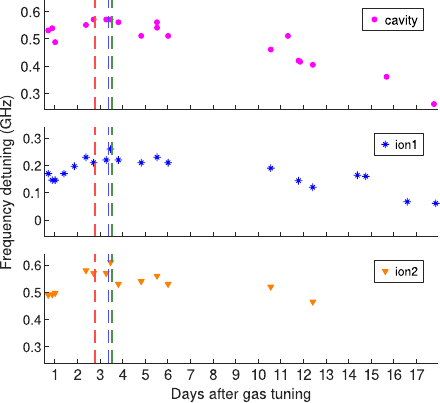}
    \caption{\textbf{Spectral wandering of the nanocavity and two ion transitions over 18 days.} 
    Detuning from the center of the inhomogeneous broadening (1537.7607~nm) as a function of days after gas tuning. 
    Red, blue, and green dashed lines indicate the first application of magnetic fields along $B_z$, $B_y$, and $B_x$, respectively. 
    Ion~1 corresponds to the one indicated in Fig.~4a of the main text, while Ion~2 corresponds to the ion marked in orange.}
    \label{figSI:spectral_wandering}
    \end{figure*}
    
    The single-ion measurements were carried out over a period of about 18 days. 
    During this time, both the cavity resonance and the ion transition frequencies drifted (Fig.~\ref{figSI:spectral_wandering}). 
    We attribute this spectral wandering to three main effects: slow drift of the wavelength meter, evolution of the nitrogen layer deposited on the cavity surface, and changes in the local environment surrounding individual ions.
    
    When examining the behavior of the cavity and two different ions, we observe a common trend: an initial increase in frequency followed by a gradual decrease. 
    Because this trend is shared by both the cavity and the ions, we attribute it primarily to long-term drift of the wavelength meter, which is used to lock the laser.
    In contrast, during the first few days after gas tuning, the cavity frequency is likely also affected by genuine tuning due to the settling and stabilization of the nitrogen layer.
    
    Interestingly, distinct shifts appear after the first application of magnetic fields along the $z$, $y$, and $x$ directions (as indicated in Fig.~\ref{figSI:spectral_wandering}). 
    After each magnetic-field cycle, the cavity resonance remains essentially unchanged, whereas the ion frequencies shift—most prominently after applying a field along the $y$ direction, where Ion~1 has its strongest $g$-tensor component. 
    Before this, Ion~1 appears as a single spectral peak (see Fig.~\ref{figSI:cavity_res}c), but after the application of the magnetic field along $y$, it remains permanently split (as in Fig.~4a of the main text), even when the nominal magnetic field is set back to zero. 
    This behavior suggests the presence of residual magnetization, possibly arising from the vector magnet or from nearby magnetic components.
    
    We also observe changes in the relative frequency spacing between the two ions over the 18 days. 
    We attribute these variations to fluctuations in the local charge environment, likely dominated by slow laser-induced charge reshuffling~\cite{Fruh2026, Bowness2025, Zhang2025}.
    
    Finally, this slow spectral diffusion may also contribute to the slight deviations from the expected linear frequency shift versus magnetic-field intensity observed in the lifetime measurements of Fig.~5c in the main text. 
    We also cannot exclude occasional loss of laser lock during these measurements, as the reference cavity used for frequency stabilization is particularly sensitive.
    
    \subsection{Purcell enhancement}
    The Purcell factor is defined as
    \begin{equation}
    \label{eq:purcell_factor}
    F_{\mathrm{P}} = \beta \cdot \eta_{\Delta} \cdot \frac{3}{4\pi^2} 
    \left( \frac{\lambda}{n} \right)^3 
    \cdot \frac{Q}{V_{\mathrm{eff}}},
    \end{equation}
    where $n = 3.48$ is the silicon refractive index, $\lambda \approx 1540$~nm is the cavity wavelength,
    $Q = 1.6 \times 10^5$ is the quality factor, and
    $\beta=0.23$~\cite{Gritsch2022} is the branching ratio of the $Y_1-Z_1$ transition.
    The factor $\eta_{\Delta}$ accounts for the spectral detuning $\Delta$ between the ion and the cavity resonance, and is defined as
    \begin{equation}
    \eta_{\Delta} = \frac{1}{1 + \left( \frac{2\Delta}{\kappa} \right)^2},
    \end{equation}
    where $\kappa/2\pi = 1.2$~GHz is the cavity linewidth.
    The effective (local) mode volume experienced by an ion located at position $\mathbf{r}_i$ is
    \begin{equation}
    V_\mathrm{eff}(\mathbf{r}_i)
    =
    V \,
    \frac{|\mathbf{E}_\text{max}|^2}
         {\eta_d \cdot |\mathbf{E}(\mathbf{r}_i)|^2},
    \end{equation}
    where $\mathbf{E}_\text{max}$ denotes the electric field amplitude at the position of maximum cavity intensity, 
    and $\eta_d$ is the dipole alignment factor defined as
    \begin{equation}
    \eta_d = 
    \frac{|\mathbf{d_\parallel}\cdot \mathbf{E}(\mathbf{r}_i)|^2}
    {|\mathbf{d_\parallel}|^2 \cdot|\mathbf{E}(\mathbf{r}_i)|^2}.
    \end{equation}
    The quantity $\eta_d$ accounts for the relative orientation between the transition dipole and the local cavity field polarization.
    
    The intrinsic mode volume of the cavity mode is given by
    \begin{equation}
    V = 
    \frac{
        \displaystyle \int n^2\,|\mathbf{E}(\mathbf{r})|^2\, d^3\mathbf{r}
    }{
        \displaystyle n^2\,|\mathbf{E}_\text{max}|^2
    }.
    \end{equation}
    
    We define the dimensionless parameter
    \begin{equation}
    V_n \equiv    
\left[V\cdot \left(\frac{n}{\lambda}\right)^3\right]\cdot 
    \frac{|\mathbf{E}_\text{max}|^2}
         {  |\mathbf{E}(\mathbf{r}_i)|^2}
    =
    \eta_d \, V_{\mathrm{eff}}\, \left(\frac{n}{\lambda}\right)^3,
    \end{equation}
    which reflects how far the ion is located from the field maximum, from a comparison with $V$.
    From the measured Purcell factor $F_\mathrm{P}$, and knowing $\eta_d$ (from the magnetic class) and $\eta_{\Delta}$, 
    we can relate $V_n$ to these quantities. 
    By rearranging Eq.~(\ref{eq:purcell_factor}), we obtain
    \begin{equation}
    V_n = \frac{3 \eta_{\Delta}\eta_d  \beta Q}{4\pi^2F_P} 
         = \frac{\eta_{\Delta}\eta_d \cdot 2.8\times 10^3}{F_P}.
    \end{equation}
    We report the corresponding parameters for three different ions in Table~\ref{tab:purcell}.  
    
    \begin{table}[htb!]
    \caption{\textbf{Summary of the measured and calculated parameters for three representative ions coupled to the cavity.}
    For each ion, we report the detuning $\Delta$ from the cavity, the corresponding detuning factor $\eta_{\Delta}$, the magnetic class, the dipole alignment factor $\eta_d$, the measured cavity-enhanced lifetime $\tau$, and the corresponding Purcell factor $F_{\mathrm{P}}$ calculated with respect to the bulk lifetime of \SI{172}{\micro\second}. 
    The derived dimensionless parameter $V_n$ quantifies the effective normalized mode volume experienced by each ion, incorporating only spatial effects. 
    By comparison with the simulated mode volume $V = 0.32(\lambda/n)^3$, larger $V_n$ values indicate that the ions are located away from the field antinode of the cavity mode and/or a deviation of the actual mode volume from the designed value.
    }
    \label{tab:purcell}
    \begin{tabular}{ccccccc}
    \toprule
    $\Delta/2\pi$~(GHz) & $\eta_{\Delta}$ & Class & $\eta_{\mathrm{d}}$ & $\tau$ (\SI{}{\micro\second}) & $F_{\mathrm{P}}$ & $V_n$ \\
    \midrule
    $-0.34$ & $0.76$ & 1 & $0.75$ & $2.44$ & $70$ & $22.5$\\
    $-0.03$ & $0.997$ & 6 & $0.58$ & $3.00$ & $57$ & $28.3$\\
    $0.19$ & $0.91$ & 2 & $0.33$ & $0.77$ & $223$ & $3.9$\\
    \botrule
    \end{tabular}
    \end{table}

    Considering that we simulate a mode volume of $V = 0.32\,(\lambda/n)^3$, 
    the experimentally extracted $V_n$ values are comparatively high.
    Such large $V_n$ indicate that the emitters do not experience the maximum cavity field intensity, 
    i.e., they are likely positioned away from the field antinode where $|\mathbf{E}(\mathbf{r}_i)| < |\mathbf{E}_\text{max}|$, 
    and/or that the fabricated cavity deviates from the ideal simulated geometry, resulting in a larger effective mode volume.
    In other words, the ions are not located at the optimal field position and/or the real cavity exhibits weaker optical confinement than predicted by simulation.
    Most likely the actual mode volume is larger than the simulated one; for instance, the simulated quality factor is $7.7 \times 10^6$, much higher than the measured one.

    % class 1, -0.34 GHz from cav (factor 0.76), F_P = 77.9, (2.44 us), eta_d = 0.75 --> x = 0.37
    % class 6, -0.03 GHz from cav (factor 0.997), F_P = 63.3 (3 us), eta_d = 0.58 --> x = 0.39, 
    % class 2, 0.19 GHz from cav (factor 0.91), F_P = 246.7 (0.77 us), eta_d = 0.33 --> x = 0.20

    \subsection{Single-ion lifetime fits}
    We elaborate on the lifetime fits of the cavity-enhanced ions discussed in Fig.~4a of the main text. 
    For a single ion, the time-dependent fluorescence follows a single-exponential decay:
    \begin{equation}\label{eq:single_exp_lifetime}
        I(t) = I_0\, e^{-(t-t_0)/\tau_0} + N,
    \end{equation}
    where $\tau_0$ is the lifetime and $N$ accounts for background noise. 
    When two distinct timescales are visible—indicating that at least two ions contribute to the detected signal—a double-exponential model is required:
    \begin{equation}\label{eq:double_exp_lifetime}
        I(t) = I_0\, e^{-(t-t_0)/\tau_0} + I_1\, e^{-(t-t_0)/\tau_1} + N,
    \end{equation}
    where $\tau_0$ and $\tau_1$ represent the two characteristic decay times.
    
\begin{figure*}[htb!]
    \centering
    \includegraphics[]{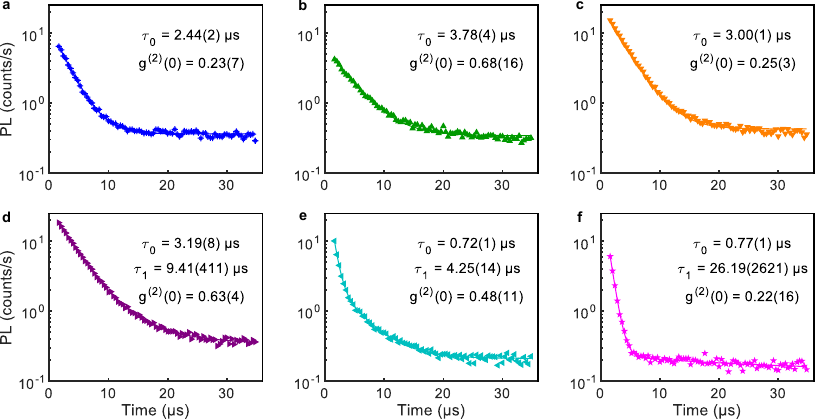}
    \caption{\textbf{Time-resolved PL from cavity-enhanced ions.} 
    The plots correspond to the insets of Fig.~4a in the main text, now shown over a time range extending to \SI{35}{\micro\second} after the end of the excitation pulse. 
    \textbf{a.} Ion 1, fitted with a single-exponential decay [Eq.~(\ref{eq:single_exp_lifetime})]. 
    \textbf{b, c.} The traces show an additional long component that is too weak to be reliably fitted; therefore, a single-exponential model is used. 
    \textbf{d, e.} Clear evidence of two decay times, requiring a double-exponential fit [Eq.~(\ref{eq:double_exp_lifetime})]; the corresponding $g^{(2)}(0)$ values confirm that multiple ions contribute to the signal. 
    \textbf{f.} A fast and a very slow component are observed, also modeled with a double exponential.}
    \label{figSI:singles_lifetime}
    \end{figure*}

    Examples of these fits are shown in Fig.~\ref{figSI:singles_lifetime}. 
    Panels~a–c are fitted with Eq.~(\ref{eq:single_exp_lifetime}), while panels~d–f require the double-exponential model of Eq.~(\ref{eq:double_exp_lifetime}). 
    Panel~a corresponds to ion 1 and exhibits the cleanest single-exponential decay, consistent with the response of an isolated emitter. 
    In panels~b and~c, a weak and slowly decaying background component is visible but too faint and extended to be reliably fitted; a single-exponential model still captures the dominant decay channel. 
    For panels~d and~e, two well-separated timescales are evident, confirming the presence of multiple ions contributing to the detected emission. 
    This interpretation is supported by their corresponding $g^{(2)}(0)$ values (see Fig.~4a in the main text), which show incomplete antibunching. 
    Panel~f displays a strong contrast between a very fast and a very slow decay component; although two emitters are present, the large difference in their coupling strengths results in $g^{(2)}(0)$ values still below 0.5.
    
    \subsection{Power sweeps with a single ion}
    % best fit parameters: I0 = 5e-3, P_sat = 1.31 \pm 0.13 nW, N = 6.76e-4
    
    \begin{figure*}[htb!]
    \centering
    \includegraphics[]{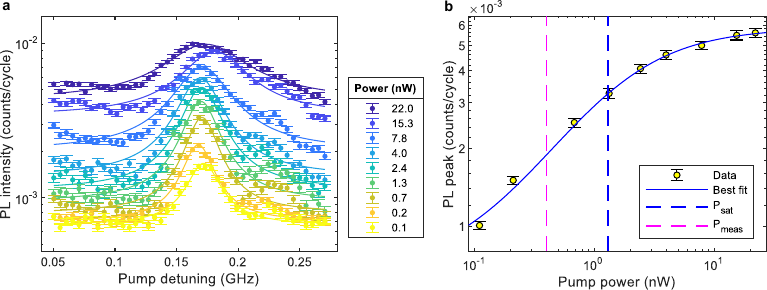}
    \caption{\textbf{Power sweeps for ion 1.}
    \textbf{a.} PL spectra measured at different excitation powers, together with Lorentzian fits. 
    \textbf{b.} Extracted PL peak intensity (after background subtraction) as a function of excitation power. 
    The solid line shows the best fit to Eq.~(\ref{eq:power_single}), while the blue dashed line marks the corresponding saturation power $P_{\mathrm{sat}} = (1.3 \pm 0.1)$~nW. 
    The magenta dashed line indicates the excitation power used for the single-ion measurements discussed in the main text.}
    \label{figSI:power_sweep_single}
    \end{figure*}
    
    Following the procedure described in Sec.~\ref{sec:power_sweeps_ensemble}, we measure the PL spectra of ion 1 as a function of the optical excitation power. 
    Each spectrum is fitted with a Lorentzian profile [Eq.~(\ref{eq:lorentzian})], as shown in Fig.~\ref{figSI:power_sweep_single}a. 
    The spectral lines exhibit slight asymmetry, which we attribute to the presence of a nearby ion that is blue-detuned from the dominant transition. 
    Spectral wandering between consecutive measurements is also observed.
    
    Analogously to the ensemble case, we extract the PL peak intensity as a function of excitation power (Fig.~\ref{figSI:power_sweep_single}b) and fit the data using a two-level saturation model:
    \begin{equation}\label{eq:power_single}
        I(P) = I_0 \left( \frac{P}{P + P_{\mathrm{sat}}} \right) + N,
    \end{equation}
    where $I_0$ is the maximum emission intensity, $P_{\mathrm{sat}}$ is the saturation power, and $N$ accounts for the residual background.
    
    In this case, the linear term included in the ensemble model (Eq.~(\ref{eq:power_ensemble})) is omitted because the experiment is carried at much lower powers and the NB waveguide is significantly shorter than the LWG, making that contribution negligible. 
    From the fit, we extract a saturation power of $P_{\mathrm{sat}} = 1.3 \pm 0.1$~nW, which is slightly higher than the optical powers used during the single-ion measurements (0.4–1~nW). 
    We choose this power regime in order to minimize power broadening while maintaining a sufficient signal-to-noise ratio.

    \subsection{$g^{(2)}(\tau)$ noise level and fit}
    % g2(0)-noise = 0.1087 +- 0.0405
    
    % a*exp(-abs(t)/tau) + 1
    % Single exponential + 1 fit results:
    % a = 0.84(3)
    % tau = 1.72(8) ms
    
    % Double exponential + 1 fit results:
    % A1*exp(-abs(t)/t1) + A2*exp(-abs(t)/t2) + 1
    % A1 = 1.34(10)
    % t1 = 0.16(2) ms
    % A2 = 0.51(3)
    % t2 = 2.88(16) ms
    
    To estimate the noise contribution to the auto-correlation function $g^{(2)}(0)$, we follow a procedure analogous to that described in Supplementary Material~VI of Ref.~\cite{Yu2023}, where the noise is treated as classical.
    The noise level is measured by closing the shutter when no light is sent to the device, thereby accounting for superconducting nanowire single-photon detector dark counts and residual background light. 
    After subtracting this noise contribution from the measured $g^{(2)}(0)$ of the selected ion, we obtain a residual value of $0.11 \pm 0.04$. 
    We attribute the remaining deviation from zero to ions that are not Purcell-enhanced within the coupling waveguide or the NB, as well as to partial spectral overlap with a nearby ion.
    
    \begin{figure*}[htb!]
    \centering
    \includegraphics[]{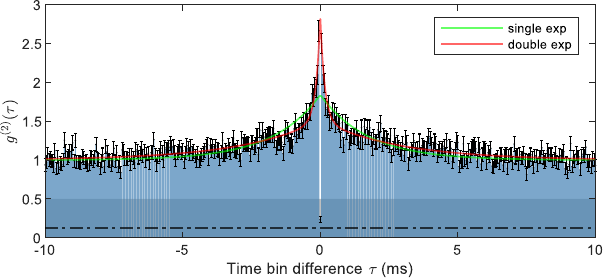}
    \caption{\textbf{Second-order correlation function $\boldsymbol{g^{(2)}(\tau)}$ for ion~1, fitted with two models.} 
    This plot is identical to Fig.~4c in the main text, but shown here over an extended time range. 
    The data are fitted with a single-exponential model (green trace) and a double-exponential model (red trace).}
    \label{figSI:g2}
    \end{figure*}
    
    The bunching behavior observed at nonzero $\tau$ in the auto-correlation function is a well-known feature in several systems~\cite{Gritsch2023, Rengstl2015, Buzzi2025, Bowness2025}, and can be described by an exponential decay:
    \begin{equation}\label{eq:g2_single}
        g^{(2)}(\tau) = A_0 e^{-|\tau|/\tau_0} + 1,
    \end{equation}
    valid for $|\tau| \geq \tau_{\mathrm{cycle}}$, with $\tau_{\mathrm{cycle}}$ the repetition period of the excitation sequence.
    
    One possible origin of this behavior is optical pumping into another Zeeman state~\cite{Gritsch2023}. 
    In our case, no external magnetic field is intentionally applied, so the spin splitting is minimal—arising only from residual stray fields—and remains within the cavity linewidth.
    We first fit the data using Eq.~(\ref{eq:g2_single}), excluding the point at $\tau = 0$, where antibunching is expected. 
    However, this single-exponential fit fails to reproduce the data accurately (see Fig.~\ref{figSI:g2}), so we repeat the fit using a double-exponential model:
    \begin{equation}\label{eq:g2_double}
        g^{(2)}(\tau) = A_0 e^{-|\tau|/\tau_0} + A_1 e^{-|\tau|/\tau_1} + 1.
    \end{equation}
    This model provides an excellent description of the experimental data.
    
    The best-fit parameters are $\tau_0 = 2.88 \pm 0.16$~ms, $A_0 = 0.51 \pm 0.03$, $\tau_1 = 0.16 \pm 0.02$~ms, and $A_1 = 1.3 \pm 0.1$. 
    Two distinct timescales are therefore identified: a fast ($\tau_1$) and a slow ($\tau_0$) component. 
    We tentatively attribute the fast component to decay into the other Zeeman state, while the slower one may arise from spectral diffusion due to laser-induced charge noise~\cite{Fruh2026, Bowness2025, Zhang2025} and nearby nuclear spins~\cite{Gritsch2025}.
    
%\clearpage


\end{document}